\documentclass[prd,twocolumn,preprintnumbers,floatfix, longbibliography]{revtex4-1}
\usepackage{amsmath}
\usepackage{amssymb}
\usepackage{bm}
\usepackage{graphicx}
\usepackage{color}
\usepackage{enumitem}
\usepackage{xcolor}
\usepackage[scr=dutchcal]{mathalpha}
\definecolor{refblue}{RGB}{0,136,187}

\usepackage{hyperref}
\hypersetup{
    colorlinks=true,
    urlcolor=refblue,
    citecolor=refblue,
    linkcolor=refblue
}
\usepackage{tikz}
\usetikzlibrary{angles,quotes,calc,arrows.meta}

\providecommand{\kh}{\hat{\bm k}}
\providecommand{\nh}{\hat{\bm n}}

\providecommand{\xGC}{\bm{x}_{\rm GC}}
\providecommand{\dx}{\Delta\bm{x}}

\providecommand{\Tr}{\mathcal{T}_r}
\providecommand{\dd}{\mathrm{d}}
\providecommand{\ee}{\mathrm{e}}
\providecommand{\ii}{\mathrm{i}}
\providecommand{\Pdy}{\mathsf{P}}
\providecommand{\Ddy}{\mathsf{D}}
\providecommand{\Sdy}{\mathsf{S}}
\providecommand{\Dop}{\mathbb{D}}
\providecommand{\Ylm}{Y_{\ell m}}

\DeclareMathAlphabet{\mathpzc}{OT1}{pzc}{m}{it}
\begin{document}
%\bibliographystyle{apsrev}
%\bibliographystyle{apsrev4-1}
%\title{Modulation of the galactic foreground by the LISA antenna pattern:\\ A fast, forward model and corrections for finite and unequal-arms}
\title{Time-Frequency Analysis for LISA: Galactic Foreground Modulation}
\author{Sudhi Mathur}
    %\email[]{sudhi.mathur@montana.edu}% Your name
    \affiliation{eXtreme Gravity Institute, Department of Physics, Montana State University, Bozeman, Montana 59717 USA}
\author{Neil J. Cornish}
    \affiliation{eXtreme Gravity Institute, Department of Physics, Montana State University, Bozeman, Montana 59717 USA}

\date{\today}
\begin{abstract}
Observations by the Laser Interferometer Space Antenna (LISA) will contain time-varying, cyclostationary confusion noise from the Galaxy, which is naturally handled using time-frequency analysis methods. We present a fast, forward model that maps the parameterized Galactic gravitational-wave luminosity density onto the annually modulated, unresolved confusion foreground in the complete set of time-delay interferometry channels. The line-of-sight luminosity integral is evaluated on a HEALPix grid and decomposed into spherical harmonics, while the detector response is decomposed analytically, independent of the Galaxy model, such that a new Galactic model costs a single map evaluation, transformation, and contraction. We compute the modulated covariance beyond the standard low-frequency limit, expanding the LISA response to second order in
finite armlength corrections, and to first order in the unequal-arm corrections. These effects change the normalized modulation curves by up to $\sim\!0.23\%$ and $0.64\%$ for finite armlength corrections at $3$ and $5$ mHz respectively, and by $\sim\!0.074\%$ and $0.71\%$ for unequal-arm corrections in the self and cross channels respectively. Using the fast, forward model and a notional disk-and-bulge Galaxy model, we construct a Fisher matrix from the time-dependent channel covariance and forecast that one year of observation constrains the disk scale width and scale height relative to the distance to the Galactic center at the $8\%$ and $12\%$ levels respectively, and leaves the strongly correlated bulge fraction and bulge radius individually unconstrained.
\end{abstract}
%\keywords{first keyword, second keyword, third keyword}
\maketitle

% ======================================================================
\section{INTRODUCTION} \label{sec:Introduction}

The Laser Interferometer Space Antenna (LISA) \cite{Amaro2017} will observe gravitational waves from a large Galactic population of compact binaries. A small fraction of these systems will be individually resolvable, while the remaining signals will overlap within each frequency bin and form an unresolved confusion foreground in the millihertz band, exceeding instrument noise at frequencies where LISA is the most sensitive \cite{Crowder2007,Littenberg2023,Nelemans2001}. The confusion foreground is anisotropic as the Galactic binary population is concentrated along the nonuniform stellar distribution of the Milky Way. The orientation of LISA's triangular constellation with respect to the foreground also changes as the constellation revolves around the Sun and rotates about its guiding center. The anisotropic foreground modulated by LISA's time-dependent antenna pattern makes the observed signal cyclostationary \cite{seto2004,Edlund:2005ye}. 

Time-frequency methods are a natural way to model such non-stationary processes \cite{Cornish:2020odn, Cornish:2026tjt, Cornish:2026mrt}. For non-stationary noise, the Fourier-domain noise covariance matrix is no longer diagonal \cite{Cornish:2020odn}, and a diagonal approximation can only represent a time-averaged noise level \cite{Cornish:2026tjt}. This approximation is inadequate for the Galactic confusion foreground, whose strong annual variation causes the signal-to-noise ratio (SNR) of a short-duration signal to vary by a factor of two over the year \cite{Cornish:2026tjt}. In contrast, the noise covariance matrix in the Wilson-Daubechies-Meyer (WDM) time-frequency basis is well approximated as diagonal for noise varying slowly across a wavelet pixel \cite{Cornish:2025awt} and captures the annual SNR modulation of short-duration signals \cite{Digman2022,Cornish:2026tjt}. Building on cyclostationary treatments of the foreground \cite{seto2004,Digman2022}, we construct the covariance matrix directly from a parameterized Galactic luminosity density, and extend it beyond the low-frequency, equal-arm limit.

Characterizing the Galactic foreground is essential for robust parameter inference in the LISA band. This includes recovering and characterizing overlapping sources \cite{Crowder2007, Littenberg2023}, separating weak stochastic backgrounds from the foreground and instrument noise \cite{Adams2014, Boileau2021}, constraining the spatial distribution of the Galactic binary population \cite{Adams2012, Korol2019, Breivik2020}, and estimating the modulation-dependent sensitivity required for reliable multi-messenger alerts \cite{Digman2022}. The unresolved Galactic foreground can be modeled as a nuisance noise component or as an astrophysical signal. As nuisance, it is a cyclostationary background whose time-dependent spectrum must be modeled jointly with instrumental noise and other gravitational wave components \cite{Adams2014, Boileau2021, Digman2022}. As a signal, its annual modulation encodes lower-order angular information about the spatial distribution of the unresolved Galactic binary population \cite{seto2004, Criswell2025, Breivik2020}. 

Existing treatments of the annual modulation either adopt a physical parameterization of the Galaxy with a computationally expensive forward model, or they use a fast model with a map-based representation or fixed angular template. Population realization forward methods simulate a synthetic Galactic binary population, propagate it through LISA's antenna response, and measure the modulation empirically \cite{Digman2022,Boileau2021}. They preserve source discreteness and population structure, at the cost of substantial computation and no analytic mapping from Galactic density parameters to the modulation. Model-agnostic map-making methods instead reconstruct the angular foreground in a pixel or spherical harmonic basis \cite{Banagiri2021}. Although this approach can recover arbitrary distributions of gravitational wave power, including angular structure not represented by a simple disk-and-bulge model, its inference variables are map coefficients rather than Galaxy density parameters, and interpreting the recovered map requires an additional, model-dependent step. Template-based stochastic analyses fit astrophysically-motivated, projected sky templates, connecting the modulation of the foreground to the spatial distribution of the Galaxy \cite{Criswell2025}. Each template must be constructed in advance and has a fixed angular distribution. Such analyses therefore test whether the sky is consistent with a given Galaxy model rather than directly sampling the Galaxy parameters. Population-synthesis calculations take the opposite route by evolving a binary population to predict the frequency distribution, luminosity function, source discreteness, selection effects, and consequently the residual foreground left after resolvable binaries are identified and subtracted \cite{Breivik2020, Korol2019, Timpano2006}. A catalog-level calculation that combines binary evolution, spatial structure, source selection, and a stochastic realization into a single foreground prediction is far too expensive to evaluate inside a likelihood, thus such predictions are used to calibrate templates rather than to explore parameters.

Within the LISA global fit, the unresolved Galactic foreground is a part of the noise model, shared by all source analyses and affecting both the likelihood and the prior \cite{Littenberg2020,Littenberg2023}. The noise covariance appears in the joint likelihood of every source model, and is refit to the residual after subtracting the resolved signals in each cycle of the sampler \cite{Littenberg2023}. The noise level also sets the amplitude prior of the resolved binaries through the expected SNR \cite{Littenberg2020}. Thus, the foreground and resolved binaries are inferred jointly. Replacing the phenomenological foreground with a physically parameterized one adds the Galactic parameters to the noise model. As the Galactic parameters are sampled, the modulation must be recomputed for every proposed set of these parameters in order to evaluate the likelihood. Thus, the cost of computing the modulation adds to the cost of each likelihood evaluation, which must remain low for the global fit to converge. This makes an accurate fast, forward model essential for characterizing the foreground, which lies, within the noise model, at the core of the global fit. Moreover, current global fit implementations model the foreground as stationary noise, and replacing this with a cyclostationary, time-frequency foreground model has been a priority \cite{Littenberg2023}. 

We present a fast, forward model that maps {\em any} parameterized model of the Galactic gravitational-wave luminosity density onto the modulation of the power in the complete set of time-delay interferometry (TDI) channels. The Galaxy luminosity density model can easily be generalized to include local over-densities from a bar or spiral arms, and the contribution from nearby satellite galaxies such as the large and small Magellanic clouds. Our analysis method is unchanged for more complicated density distributions.

The gravitational-wave luminosity density is integrated along the line of sight on a HEALPix grid \cite{Gorski2005} and decomposed into spherical harmonics, while the detector response is computed analytically once and tabulated independently of the Galaxy. This separation requires the angular distribution of power to be independent of frequency, which is standard in anisotropic background analyses \cite{RomanoCornish2017,seto2004}. A new Galaxy model then costs a single sky map evaluation, transformation, and contraction. Our model complements population synthesis methods, as it applies to any gravitational-wave luminosity density that can be integrated along the line of sight, including densities constructed directly from catalogs. For a specified spatial luminosity density, our model can isolate the geometric contribution to the modulated covariance and provide a rapid framework for hyperparameter inference.

Extending the unresolved Galactic foreground model to include finite and unequal-arm corrections is also important for unbiased recovery across the global fit. As the modulated foreground enters the covariance matrix common across every module of the global fit \cite{Littenberg2023}, modeling errors in its angular or temporal structure can be absorbed into the foreground spectral parameters, instrument noise, or the resolved binary parameters at the confusion boundary. We quantify the resulting finite and unequal-arm effects in the TDI channel modulation curves and determine the frequencies at which the corrections become significant. 

Our main developments are:
\begin{enumerate}[itemsep=2pt, topsep=3pt]
    \item A fast, forward model for a parameterized Galaxy~that
    \begin{itemize}[label=$\diamond$, leftmargin=1em, itemsep=1pt, topsep=0pt]
        \item Maps an arbitrary Galactic luminosity density directly to the time-dependent TDI channel foreground covariance.
        \item Separates the Galactic model from the detector response. Rapidly evaluates modulation for new Galaxy density models without re-synthesizing populations or detector response.
        \item Enables computationally efficient exploration of Galactic parameter space and Fisher forecasts for information carried by the annual modulation.
    \end{itemize}
    \item Finite and unequal-arm corrections applied to the modulated foreground, where we
    \begin{itemize}[label=$\diamond$, leftmargin=1em, itemsep=1pt, topsep=0pt]
        \item Extend the finite-arm treatment from the self-correlation TDI channels \cite{Cornish2002} to the full set of self and cross-correlation channels and retain the full complex covariance.
        \item Calculate the leading unequal-arm corrections to the modulation curves.
        \item Quantify correction effects on the modulation curves and identify the frequencies at which these corrections become important.
    \end{itemize}
\end{enumerate}

The paper is structured as follows. Sec.~\ref{Sec:Methods} separates the Galaxy luminosity integral from the spherical harmonic decomposition of the detector response and defines their contraction. The spherical harmonic coefficients of the leading-order self and cross-correlation detector response, along with the finite-arm and unequal-arm corrections, are then computed in the detector frame and Wigner-rotated to the barycentric frame~\cite{Cornish2002}. HEALPix integral of the line-of-sight luminosity function and the Fisher matrix on the Galactic parameters are further described. Sec.~\ref{sec:Results} plots the scaled self and cross-channel modulation curves of the unresolved foreground at the leading-order, and for the finite and unequal-arm corrections. The percentage corrections introduced by the finite-arm effects are plotted for frequencies $3, 5$ and $10 \,\text{mHz}$. Frequency dependence of the percentage correction to the scaled leading-order response due to the finite-arm effects is quantified and plotted. The modulation contributions of the spherical harmonic $\ell$ modes and the fractional power in the annual modulation harmonic $k$ modes for the scaled, self and cross leading-order modulation curves, and finite-arm and unequal-arm corrections are also highlighted. The sensitivity of the modulation curves to each Galaxy parameter and the change in finite and unequal-arm corrections across astrophysically motivated Galaxy parameter ranges are also presented. The Fisher forecast for the notional Galaxy is shown as a corner plot. Sec.~\ref{sec:Conclusion} summarizes our conclusions. Sec.~\ref{sec:future} discusses broader implications of our work and future directions, including extending to flexing arms and second-generation TDI, incorporating more realistic Galactic structure, and potentially performing joint inference on the resolved and unresolved Galactic binary populations.

\section{Methods}\label{Sec:Methods}

Averaging over the random phases, polarization angles, inclinations, and masses of the unresolved binary population reduces the LISA channel cross-spectrum to a sky integral of the Galactic luminosity weighted by the polarization-summed detector response. Expanding the luminosity and response in spherical harmonics, orthonormality then reduces the sky integral to a time-dependent contraction
\begin{equation}
\begin{aligned}
\left\langle X_j X_{k}^{*}\right\rangle(t)
& \;\propto\;
\int \dd\Omega\; \mathcal{L}(\theta,\phi)\, R^{jk}(f,\kh,t) \\
& \;\propto\;
\sum_{\ell=0}^{\ell_{\max}}\sum_{m=-\ell}^{\ell}
a^*_{\ell m}\,b^{jk}_{\ell m}(t).
\end{aligned}
\label{eq:factorization}
\end{equation}
For an unpolarized background with no correlations between distinct frequencies, polarizations, or sky directions, the delta functions \(\delta_{AA'}\delta(f-f')\delta^2(\hat{k}-\hat{k}')\) reduce the double sky integral and the double polarization sum \cite{Cornish2001, Cornish2002, AllenOttewill1997, Ungarelli2001}. The $a_{\ell m}$ are the spherical harmonic coefficients of the line-of-sight-integrated gravitational-wave luminosity given as
\begin{equation}
\begin{aligned}
& \mathcal{L}(\theta,\phi)\propto
\int_0^{\infty}\rho(\mathscr{r},\theta,\phi)\,\dd \mathscr{r},
\\
& a_{\ell m}=
\int \dd\Omega\;
\mathcal{L}(\theta,\phi)\,\Ylm^*(\theta,\phi),
\end{aligned}
\label{eq:alm}
\end{equation}
where $\mathscr{r}$ is the heliocentric distance along the line of sight. The \(b_{\ell m}^{jk}(t)\) are the spherical harmonic coefficients of the polarization and inclination-averaged $jk$ channel response  
\begin{equation}
\begin{aligned}
    & R^{jk} = \sum_{A=+,\times} F^{A}_{j}(f,\kh,t)\,F^{A*}_{k}(f,\kh,t), \\
     & b^{jk}_{\ell m} = \int \dd\Omega R^{jk}\,\Ylm^\ast,
\end{aligned}\label{eq:blm&Rjk}
\end{equation}
where $F^{A}_{j}$ is the antenna pattern of channel $j$ for polarization $A \in \{+,\times\}$. The frequency-dependent, population-averaged source luminosity is carried by the foreground spectral normalization, while the $a_{\ell m}$ describe only its frequency-independent angular distribution \cite{Cornish2001, Cornish2002}. The brackets in Eq.~\eqref{eq:factorization} denote the time-average over the unresolved population, while LISA's orbital motion gives the explicit time dependence of $b^{jk}_{\ell m}(t)$ and generates the annual modulation. The detector response is computed once and tabulated in terms of the detector frame spherical harmonic coefficients. These are then rotated into the barycentric frame by time-dependent Wigner rotation matrices. Thus, evaluating a new Galactic density model requires only the sky map evaluation, transformation to the Galactic coefficients $a_{\ell m}$, and contraction with the precomputed response coefficients $b_{\ell m}^{jk}(t)$ in Eq.~\eqref{eq:factorization}.

We extend the $\mathcal{O}(f/f_\ast)^2$ finite-arm corrected $b^{jk}_{\ell m}(t)$ from the self-correlation channels \cite{Cornish2002, Cornish2001} to the cross-correlation channels, where \(f_\ast \equiv 1/(2\pi L)\) is the transfer frequency along each arm, and calculate the leading-order unequal-arm corrections at $\mathcal{O}(\epsilon)$. The full frequency-dependent LISA response is well established \cite{CornishRubbo2003} and is currently employed in the global fit, under the approximation of a rigid, equal-arm constellation. It enters the waveforms of individually resolved sources, the TDI transfer functions of the instrument noise model \cite{Littenberg2023}, and the channel covariances used in stochastic-background reconstructions \cite{RosatiLittenberg2025} and is retained at all orders in $f/f_\ast$. The annual modulation of the Galactic confusion foreground, by contrast, is modeled to leading-order in the low-frequency limit \(f \ll f_\ast\) \cite{seto2004}. A LISA constellation with \(L = 2.5 \times 10^9\,\mathrm{m}\) gives \(f_\ast \simeq 19.1\,\mathrm{mHz}\) \cite{Amaro2017}. While the foreground is strongest at frequencies well below $f_\ast$, it remains present into the several-millihertz range, where $f/f_\ast$ is no longer parametrically small, for example $f/f_\ast \simeq 0.16$ at $3\,\mathrm{mHz}$, $0.26$ at $5\,\mathrm{mHz}$, and $0.52$ at $10\,\mathrm{mHz}$. Thus, the low frequency assumption becomes increasingly inaccurate in the upper part of the foreground-dominant band. We quantify the impact of these finite-arm corrections on the modulated Galactic foreground covariance. 

We also account for the leading-order corrections in the unequal-arm limit. The LISA constellation is not equilateral. At leading-order, the unequal-arm correction changes the opening angles of the constellation and arm-product weights of the TDI response. Writing the arm lengths as $\ell_{ij} = L(1 + \epsilon_{ij})$, the leading-order static unequal-arm effect is set by $\epsilon_{ij}$, which vary over the orbit and add annual sidebands to the modulation covariance. We treat finite-arm and static unequal-arm effects as separate perturbations, neglecting mixed terms of order $\epsilon f/f_\ast$ or $\epsilon (f/f_\ast)^2$.

The $A$, $E$, $T$ basis is signal-orthogonal only for an isotropic sky \cite{Hartwig2023}. At leading-order, $\langle AE^{*}\rangle$ is a pure $\ell = 4$, $m = \pm 4$ response (Sec.~\ref{subsubsec:aet}, Fig.~\ref{fig:aet}), thus an anisotropic foreground populates it even in the equal arm length limit, and unequal arms break the remaining structure independently. Thus, we work with the full $3\times3$ complex covariance in the $X$, $Y$, $Z$ basis.

We adopt a notional two-component disk-and-bulge model described by five parameters, including the bulge fraction and radius, disk width and height, and distance to the Galactic center \cite[Eq. 13]{Adams2014}. Although this simplistic model omits features such as the Galactic bar, spiral arms, and a distinct thick disk, it provides a smooth, low-dimensional parameter space which we use to quantify information carried by the foreground modulation. The foreground anisotropy provides access to spatial information that is largely absent from the sky-averaged spectrum, and LISA's time-varying antenna pattern imprints this anisotropy as an annual modulation of the channel covariance \cite{seto2004}. Using our fast, forward model, we compute the modulated TDI channel covariance across this parameter space and construct a Fisher matrix for the Galactic parameters. We then identify which parameters the modulation constrains and forecast their measurement precision.

The resolved binary and unresolved foreground can provide complementary constraints on the parameters of a common Galactic model. Individually resolved binaries have previously been used to infer Galactic population hyperparameters \cite{Adams2012,Korol2019}. In hierarchical Bayesian analyses of the resolved population, catalogs of roughly \(5\times10^3\) short-period binaries constrain the chirp-mass distribution and a characteristic Galactic disk width at the few-percent level \cite{Adams2012}, while forecasts for the larger population expected over a full LISA mission constrain scale-length parameters of the Galactic baryonic components at the few-percent to \(10\%\) level \cite{Korol2019}. While these analyses obtain spatial information from individually resolved systems, the unresolved binary population depends on the same underlying Galactic distribution. For the unresolved white-dwarf foreground, the sky-averaged spectrum is primarily sensitive to population normalization rather than spatial geometry \cite{Benacquista2006,Georgousi2023}. Changes in disk scale height affect the confusion foreground amplitude and individual source resolvability, with the constraints depending on whether the population is normalized at fixed local density or fixed total number of binaries \cite{Benacquista2006}. Models with disk scale heights \(Z_d \lesssim 600\,{\rm pc}\) at fixed total stellar mass are difficult to distinguish using the spectral shape of the stochastic foreground alone, while the foreground amplitude and spectral model constrain the total Galactic stellar mass at the percent level \cite{Georgousi2023}. By contrast, the foreground anisotropy retains geometric information about the source distribution and has been used to further constrain Galactic structure \cite{Breivik2020,Criswell2025}. For example, the spherical harmonic power spectrum of population synthesis foreground maps has been used to forecast constraints on the vertical scale height of the Galactic binary population at the \(50\)--\(200\,{\rm pc}\) level, depending on the assumed angular resolution \cite{Breivik2020}. We retain this anisotropy information in the modulated covariance and use our fast, forward model to constrain a continuously parameterized Galactic density model.

Sec.~\ref{subsec:leading} derives the arm-weighted TDI channels and the general detector response in the frame of the constellation. Sec.~\ref{subsec:lo_coeff} computes the spherical harmonic coefficients of the leading-order detector response, while Sec.~\ref{SubSec:FiniteArm} carries the frequency expansion to second order and Sec.~\ref{subsec:unequal} relaxes the equilateral assumption. Sec.~\ref{subsec:Wigner} restores the annual time dependence by a Wigner rotation to the barycentric frame. Secs.~\ref{subsec:healpix} and \ref{subsec:fisher} describe the sky map evaluation, the transformation to the Galactic coefficients, and the Fisher analysis.

\subsection{Detector frame response}\label{subsec:leading}

\subsubsection{Why the detector frame}
The spherical harmonic coefficients of the detector response channels \(b_{\ell m}^{jk}(f, t)\) can be computed in any frame. In the barycentric frame the detector arm vectors $\nh_{ij} = \nh_{ij}(\alpha,\lambda)$ carry the orbital phase of the guiding center $\alpha = 2 \pi f_m t + \kappa$ and the relative phases of the spacecrafts within the constellation $\beta = 2 \pi n/3 + \lambda$ where $(n = 0,1,2)$ for spacecrafts 1, 2 and 3, $f_m = 1\,\mathrm{yr}^{-1}$, and parameters $\kappa$ and $\lambda$ give the initial ecliptic longitude and orientation of the constellation. Thus,
\begin{equation}
b^{jk}_{\ell m}(f,t) = \int \dd\Omega\;
R^{jk}\!\left(f,\kh;\nh(\alpha,\lambda)\right)\Ylm^*
\label{eq:blm-bary}
\end{equation}
carries $(\alpha,\lambda)$ explicitly through the integrand.
In the frame of the constellation, by contrast, the arms are fixed, $\partial_t\,\nh_{ij} = 0$. The integrand depends only on
$(\bar\theta,\bar\phi,f)$,
\begin{equation}
\bar b^{\,jk}_{\ell m}(f) = \int \dd\bar\Omega\;
\bar R^{jk}(f,\bar\theta,\bar\phi)\,\Ylm^*(\bar\theta,\bar\phi),
\label{eq:blm-det}
\end{equation}
and no orbital parameter is carried through the integrand. The resulting detector frame $\bar b^{\,jk}_{\ell m}(f)$ are simpler expressions than their barycentric frame counterparts $b^{jk}_{\ell m}(f,t)$, and all of their annual time dependence is restored by a single Wigner rotation from the detector to the barycentric frame \cite{Cornish2002}.

We place spacecraft 1 at the origin, and spacecraft 2 and 3 in the $\bar x\bar y$ plane, with unit arm vectors 
\begin{equation}
\nh_{ij} = (\cos{\Psi_{ij}}, \sin{\Psi_{ij}}, 0)
\label{eq:nij}
\end{equation} 
where $\Psi_{ij}$ are the azimuthal angles in the detector frame. 

\subsubsection{The Michelson variable with arm-length weights}
\label{subsec:tdi}
The first-generation TDI Michelson variable groups six one-way measurements into round trips $\Lambda_{ij}$ measured at spacecraft $i$. The far laser cancels within each round trip, while the local laser survives and is multiplied by the round-trip delay along the arm,
\begin{equation}
\Lambda_{ij} = -\left(1-\Dop_{iji}\right)C_i + 2\pi\nu_0 \mathcal{R}_{ij},
\label{eq:roundtrip}
\end{equation}
where the delay operator $\Dop_{ij}\mathcal{F}(t) = \mathcal{F}(t-\ell_{ij})$, phase noise for laser on spacecraft $i$ pointed towards spacecrafts $j$ and $k$ is $C_{ji}(t) \simeq C_{ki}(t) \equiv C_{i}(t)$, the angular frequency of the laser is $ 2 \pi \nu_0$, and the round-trip path-length variation can be written as
\begin{equation}
    \mathcal{R}_{ij} = \delta \ell_{ij} (t-2 \ell_{ij}) + \delta \ell_{ji}(t-\ell_{ji}),
\label{eq:path-length-variation-RT}
\end{equation}
where $\delta\ell_{ij}(t)$ is the one-way path-length variation for a photon emitted by spacecraft $i$ at time $t$ and received at spacecraft $j$.
The four-link difference $\Lambda_{ij}-\Lambda_{ik}$ leaves a laser residual $\left(\Dop_{iji}-\Dop_{iki}\right)C_i (t)$ that vanishes only for equal arms. The combination that cancels it for unequal but constant arms and gives the TDI-1 channels is
\begin{equation}
X_i(t) \equiv \left(1-\Dop_{iki}\right)\Lambda_{ij} - \left(1-\Dop_{iji}\right)\Lambda_{ik},
\label{eq:X}
\end{equation}
with $i,j,k$ following cyclic permutation \cite{Armstrong1999,Tinto2002,Dhurandhar2002,TintoDhurandhar2021}. The delay operators commute for unequal but constant arms with $\ell_{ij} = \ell_{ji}$, canceling the laser residual completely. Fourier transformation over a window much larger than the wave period $1/f$ and light-crossing time $L$, but much smaller than $1/f_m$ gives
\begin{equation}
X_i(t) = \int \dd f\; \tilde X_i(f;t)\,\ee^{2\pi \ii f t},
\label{eq:window}
\end{equation}
where $\tilde X_i(f;t)$ is the slowly varying spectrum. Under the Fourier transform the delay operators contribute phase shifts such that $\widehat{\Dop}_{ij}(f) = \ee^{-2\pi \ii f \ell_{ij}}$ and $\widehat{\Dop}_{iji}(f) = \ee^{-4\pi \ii f \ell_{ij}}$. The one-way path-length variation along spacecrafts $i\to j$ can be written in the frequency domain~\cite{Rubbo2004}~as
\begin{equation}
\begin{aligned}
\delta\ell_{ij}(t) &= \ell_{ij}\int \dd f\;\tfrac{1}{2}\,\Pdy_{ij}\,\mathcal{T}(\mu_{ij})
  : \tilde{\bm h}(f)\,\ee^{2\pi\ii f\xi_i(t)},\\
\mathcal{T}(\mu_{ij}) &= \mathrm{sinc}\!\left[\tfrac{\chi_{ij}}{2}(1-\mu_{ij})\right]
  \ee^{\ii\chi_{ij}(1-\mu_{ij})/2},
\end{aligned}
\label{eq:oneway}
\end{equation}
where $\Pdy_{ij}\equiv\nh_{ij}\otimes\nh_{ij}$ is the arm dyad, $\tilde{\bm h}(f)$ is the Fourier transform of the gravitational-wave tensor ${\bm h}(\xi)$ in the transverse-traceless gauge, $\xi_i=t-\kh\cdot\bm x_i$ is the retarded time at the emitting spacecraft, $\kh$ is the unit vector in the gravitational-wave propagation direction, $\bm x$ is the spacecraft position vector, $\mathcal{T}(\mu_{ij})$ is the one-way transfer function, where $\mu_{ij}\equiv\kh\cdot\nh_{ij}$ and $\chi_{ij}\equiv f/f_{\ast\,ij}$ for transfer frequency $f_{\ast\,ij}\equiv 1/(2\pi\ell_{ij})$. The colon denotes a double contraction, ${\bf a}:{\bf b} = a^{ij}b_{ij}$. As the wave variable $\xi$ is origin-dependent,  we define $\bm{x}_n = \xGC + \dx_n$ and $A_n \equiv \kh\cdot\dx_n/L$ with respect to the guiding center. This gives
\begin{equation}
\xi_n = t_{\rm GC} - L A_n ,
\qquad
t_{\rm GC} \equiv t - \kh\cdot\xGC .
\label{eq:xiGC}
\end{equation}
In Eq.~\eqref{eq:path-length-variation-RT}, the outbound photon leaves $i$ at $t-2\ell_{ij}$ and the return photon leaves $j$ at $t-\ell_{ij}$, with $\mu_{ji}=-\mu_{ij}$. Using Eqs.(~\ref{eq:oneway}-  \ref{eq:xiGC}) and the relation $A_j-A_i=(\ell_{ij}/L)\,\kh\cdot\nh_{ij}=(1+\epsilon_{ij})\,\mu_{ij}$, the phase of each leg, given by the wave phase at emission plus the phase of the one-way transfer function, is
\begin{equation}
\begin{aligned}
&2\pi f\,\xi_i(t-2\ell_{ij})+\tfrac{\chi_{ij}}{2}(1-\mu_{ij})\\
&\qquad \qquad= 2\pi f\,t_{\rm GC}-\chi A_i-\tfrac{\chi_{ij}}{2}(3+\mu_{ij}),\\[4pt]
&2\pi f\,\xi_j(t-\ell_{ij})+\tfrac{\chi_{ij}}{2}(1+\mu_{ij})\\
&\qquad \qquad= 2\pi f\,t_{\rm GC}-\chi A_i-\tfrac{\chi_{ij}}{2}(1+\mu_{ij}),
\end{aligned}
\label{eq:rt-phases}
\end{equation}
for the outbound and return legs, respectively, where \mbox{$\chi\equiv f/f_\ast$}. Both legs share the vertex phase $\ee^{-\ii\chi A_i}$, which factors out of the round trip, while the common factor $\ee^{2\pi\ii f t_{\rm GC}}=\ee^{2\pi\ii f t}\,\ee^{-2\pi\ii f\kh\cdot\xGC}$ carries the phase $\ee^{2\pi\ii f t}$ of the Fourier transform in Eq.~\eqref{eq:window}. Thus, Eq.~\eqref{eq:path-length-variation-RT} can be written as
\begin{equation}
\mathcal{R}_{ij}(f,\kh) = \ell_{ij}\,\Pdy_{ij} : \tilde{\bm h}(f)\;
\Tr(\mu_{ij})\;\ee^{-2\pi \ii f \kh \cdot \xGC}\,\ee^{-\ii\chi A_i},
\label{eq:roundtripsignal}
\end{equation}
where the round-trip transfer function $\Tr(\mu_{ij})$ is
\begin{align}
\Tr(\mu_{ij}) = \tfrac{1}{2}\Big[&
\mathrm{sinc}\!\left[\frac{\chi_{ij}}{2}(1-\mu_{ij})\right]\ee^{-\ii\chi_{ij}(3+\mu_{ij})/2}
\nonumber\\
&+\mathrm{sinc}\!\left[\frac{\chi_{ij}}{2}(1+\mu_{ij})\right]\ee^{-\ii\chi_{ij}(1+\mu_{ij})/2}
\Big],
\label{eq:Tr}
\end{align}
and $\chi_{ij} = \chi(1+\epsilon_{ij})$. We substitute Eq.~\eqref{eq:roundtripsignal} into Eq.~\eqref{eq:X} using the definition of $\Lambda_{ij}$ in Eq.~\eqref{eq:roundtrip} and define \mbox{$\Theta_{n} \equiv \ee^{-2\pi \ii f \kh \cdot \xGC}\ee^{-\ii\chi A_n}$}. This gives
\begin{equation}
\begin{aligned}
\tilde X_i(f;t) = 2\pi & \nu_0\Big[
\left(1-\ee^{-4\pi \ii f\ell_{ik}}\right)\ell_{ij}\,\Pdy_{ij}\Tr(\mu_{ij})\\
&-\left(1-\ee^{-4\pi \ii f\ell_{ij}}\right) \ell_{ik}\,\Pdy_{ik}\Tr(\mu_{ik})
\Big] : \tilde{\bm h}(f) \;\Theta_{i} \; .
\label{eq:Xassembled}
\end{aligned}
\end{equation}
At leading-order in $\chi_{ij}$, $1-\ee^{-4\pi \ii f\ell_{ij}}\to 4\pi \ii f\ell_{ij}$, so both terms acquire the
same product $\ell_{ij}\ell_{ik}$ which factors out such that
\begin{equation}
\tilde{X}_i(f;t) = \mathcal{C}_0\, w_{X_i} \left[\Pdy_{ij}\Tr(\mu_{ij}) - \Pdy_{ik}\Tr(\mu_{ik})\right]
    :\tilde{\bm h}(f)\;\Theta_i,
\label{eq:Xfactored}
\end{equation}
where
\begin{equation}
\mathcal{C}_0(f) = 8 \pi^2 \ii \nu_0 f L^2,
\qquad
w_{X_i} = \frac{\ell_{ij}\ell_{ik}}{L^2}.
\label{eq:C0}
\end{equation}
For equal arms the weights $w_{X_i} \rightarrow 1$, but for unequal arms they are channel-dependent and to $\mathcal{O}(\epsilon)$ we can write them as $w_{X_i} = 1 + \epsilon_{ij} + \epsilon_{ik}$. 

Restoring the sky integral and projecting onto polarizations,
$h_{ab}(f,\kh) = \sum_A h_A(f,\kh)\,\textbf{e}^A_{ab}$,
\begin{equation}
\begin{aligned}
\tilde{X}_i(f;t) = \mathcal{C}_0\, w_{X_i} \sum_A & \int  \dd\Omega\;
h_A(f,\kh)\,\Theta_i \\
& \times\left[\Pdy_{ij}\Tr(\mu_{ij}) - \Pdy_{ik}\Tr(\mu_{ik})\right]
:\textbf{e}^A(\kh) ,
\end{aligned}
\end{equation}
where $\textbf{e}^A$ are the polarization tensors. We can then define 
\begin{equation}
\begin{aligned}
    & \tilde{X}_i(f;t) \equiv 2\;\mathcal{C}_0 \sum_A \int  \dd\Omega\; h_A(f,\kh) F_i^A(f, \kh), \\
    & F_i^A(f, \kh) \equiv \; \Ddy_i^{ab}(f, \kh)\; \textbf{e}^A_{ab},
\label{eq:X(f;t)|F}
\end{aligned}
\end{equation}
with detector tensors $\Ddy_i^{ab}(f, \kh)$. Thus, we can write the detector antenna pattern as
\begin{align}
F^A_i(f;\kh) = \frac{1}{2} w_{X_i} \Theta_i \left[\Pdy_{ij}\Tr(\mu_{ij}) - \Pdy_{ik}\Tr(\mu_{ik})\right]
:\textbf{e}^A(\kh),
\label{eq:anntenna}
\end{align}
and use it to calculate the spherical harmonic coefficients $b^{jk}_{\ell m}$ of the polarization and inclination-averaged channel response $R^{jk}$ in Eq.~\eqref{eq:blm&Rjk} at the leading-order, and for the finite and unequal-arm corrections. From Eqs.~\eqref{eq:X(f;t)|F} and \eqref{eq:anntenna} we can also write the complex detector tensor as
\begin{equation}
\Ddy_i(f;\kh) = \frac{1}{2} w_{X_i} \Theta_i \left[\Pdy_{ij}\Tr(\mu_{ij}) - \Pdy_{ik}\Tr(\mu_{ik})\right],
\label{eq:detector-tensor}
\end{equation}
which gives
\begin{equation}
    \Ddy_i + \Ddy_j + \Ddy_k = \mathcal{O}(\epsilon) + \mathcal{O}(\chi^3).
    \label{eq:detector-tensor-resid}
\end{equation}

\subsubsection{Reference point and the surviving phase}
The single-arm response must be phase-referenced to some point on the constellation. While referencing the single-arm response to the corresponding spacecraft is admissible for self-correlation channels \cite{Cornish2002}, it does not account for phase shifts introduced in the cross-correlation channels. We reference all single-arm responses to the centroid, $\bm{x}_n = \xGC + \dx_n$ with $A_n \equiv \kh\cdot\dx_n/L$, and account for the resulting phase factor $\ee^{-\ii\chi A_i}$ from Eq.~\eqref{eq:roundtripsignal} in Eq.~\eqref{eq:Xfactored}. For the self-correlation channels $i=j$ the phase factor is exactly unity \cite{Cornish2002}. However, for cross-correlation channels $i\neq j$ it depends only on the gauge-invariant physical difference
\begin{equation}
A_j - A_i = \frac{\kh\cdot(\bm{x}_j-\bm{x}_i)}{L}
= \left(1+\epsilon_{ij}\right)\mu_{ij},
\label{eq:Adiff}
\end{equation}
and not on the individual $A_n$. The surviving phase factor produces purely imaginary odd-$\ell$ terms at $\mathcal{O}(f/f_\ast)$ in the cross-correlation channels which do not affect the modulation curves, but should be accounted for in the quadrature spectrum and likelihood evaluations. 

\subsubsection{Cyclic and cross-to-self channel relations}
The correlation coefficients follow certain cyclic relations.  Let $\mathcal{S}_n$ be the rotation about $\bar z$ through an angle $\Phi_n$ that maps the arm set onto itself and carries channel $0$ onto channel $n$. The dyads rotate as tensors $\Pdy \to \mathcal{S}\,\Pdy\,\mathcal{S}^{\!\top}$, while the scalars follow from $\kh\cdot\mathcal{S}\bm{a} = \kh'\cdot\bm{a}$, so $\mu_{ij}$ and $A_n$ are evaluated at the rotated sky direction $\kh'$. The polarization sum is a trace over the polarization space and is invariant under the basis rotation that $\mathcal{S}_n$ induces, so the mixing of $+$ and $\times$ drops out and $R^{nn}(\kh) = R^{00}(\mathcal{S}_n^{-1}\kh)$. Substituting $\kh' = \mathcal{S}_n^{-1}\kh$ with $\dd\Omega$ invariant, and using $\Ylm(\theta,\phi+\Phi_n) = \ee^{\ii m\Phi_n}\Ylm(\theta,\phi)$ for a rotation about $\bar z$,
\begin{equation}
\begin{aligned}
& \bar b^{\,nn}_{\ell m}\!\left[\bm{\ell}\right] = \ee^{-\ii m \Phi_n}\, \bar b^{\,00}_{\ell m}\!\left[\sigma^n\bm{\ell}\right], \\
& \bar b^{\,n(n+1)}_{\ell m}\!\left[\bm{\ell}\right] = \ee^{-\ii m \Phi_n}\, \bar b^{\,01}_{\ell m}\!\left[\sigma^n\bm{\ell}\right]
\end{aligned}
\label{eq:cyclicgen}
\end{equation}
where $\bm{\ell} = (\ell_{12},\ell_{13},\ell_{23})$ denotes the constellation configuration and $\sigma^n\bm{\ell}$ the constellation with its spacecraft labels advanced cyclically $n$ times $1\!\to\!2\!\to\!3\!\to\!1$, and the ($-m$) modes are given by
\begin{equation}
    \bar b^{\,10}_{\ell m} = (-1)^m {(\bar b^{\,01}_{\ell, -m})}^\ast
\label{eq:blm-conjugate}
\end{equation}
In the equal-arm limit $\sigma^n\bm{\ell} = \bm{\ell}$, the equilateral configuration gives $\Phi_n = 2\pi n/3$ and the rotation phase becomes $\omega \equiv \ee^{-2\pi \ii m/3}$. This gives the equal-arm cyclic relations
\begin{equation}
\begin{aligned}
& \bar b^{\,YY}_{\ell m} = \omega\,\bar b^{\,XX}_{\ell m},
\qquad
\bar b^{\,ZZ}_{\ell m} = \omega^2\,\bar b^{\,XX}_{\ell m} . \\
& \bar b^{\,YZ}_{\ell m} = \omega\,\bar b^{\,XY}_{\ell m},
\qquad
\bar b^{\,ZX}_{\ell m} = \omega^2\,\bar b^{\,XY}_{\ell m} . 
\end{aligned}
\label{eq:cyclic}
\end{equation}
For unequal arms, $\Phi_n = 2\pi n/3 + \delta\Phi_n$ and the law of cosines gives $\delta\Phi_i = (\epsilon_{ij}+\epsilon_{jk}-2\epsilon_{ik})/\sqrt{3}$ which is $\mathcal{O}(\epsilon)$. Thus, the three self and cross-correlation channels in the unequal-arm limit are related by cyclic permutation of the arm length indices together with the corresponding phase factor. It is therefore sufficient to compute the seeds $\bar b^{\,00}_{\ell m}$ and $\bar b^{\,01}_{\ell m}$ as a function of the three arm lengths. The remaining correlation channel coefficients follow by permuting the arguments and applying the appropriate phase shifts.

The cross-correlation channels coefficients are also related to the self-correlation coefficients. As $F^A_n = \Ddy_n : \bm{e}^A$ and $R^{jk} = \sum_A F^A_j F^{A*}_k$, we can define the sesquilinear function
\begin{equation}
\mathcal{B}_{\ell m}\!\left[\bm{U},\bm{V}\right] \equiv \int \dd\Omega\,
\sum_A \left(\bm{U}\!:\!\bm{e}^A\right)\left(\bm{V}\!:\!\bm{e}^A\right)^{*}
\Ylm^{*}(\kh),
\label{eq:sesqui}
\end{equation}
such that 
\begin{equation}
    \bar b^{\,jk}_{\ell m} = \mathcal{N}\, \mathcal{B}_{\ell m}[\Ddy_j,\Ddy_k],
    \label{eq:Blm{Di,Dj]}}
\end{equation} 
which we use to derive the relationship between self and cross-correlation coefficients in Sec. \ref{subsec:lo_coeff}.

\subsection{Leading-order coefficients}\label{subsec:lo_coeff}

In the equal-arm length limit $\epsilon_{ij} \rightarrow 0$, placing spacecraft 1 at the origin, spacecraft 2 along $\bar x$, and spacecraft 3 in the $\bar x\bar y$ plane, the unit vectors defined in Eq.~\eqref{eq:nij} are
\begin{equation}
\nh_{12} = (1,0,0),\quad
\nh_{13} = \left(\tfrac12,\tfrac{\sqrt3}{2},0\right),\quad
\nh_{23} = \left(-\tfrac12,\tfrac{\sqrt3}{2},0\right),
\label{eq:arms}
\end{equation}
corresponding to the equilateral detector azimuthal angles $\Psi_{12} = 0$, $\Psi_{13} = \pi/3$, and $\Psi_{23} = 2\pi/3$, and opening angle $\arccos{(\nh_{12}\cdot\nh_{13})} = \pi/3$.

At leading-order in $\chi = f/f_\ast \rightarrow 0$ and in the equal-arm limit $\epsilon_{ij} \rightarrow 0$, the transfer function $Tr(\mu) \rightarrow 1$ and  Eq.~\eqref{eq:anntenna} reduces to 
\begin{align}
F^A_i(f;\kh) = \frac{1}{2} \Theta_i \left[\Pdy_{ij} - \Pdy_{ik}\right]
:\textbf{e}^A(\kh),
\label{eq:lo_anntenna}
\end{align}
and using $R^{jk} = \sum_{A} F^{A}_{j}\,F^{A*}_{k}$, we get the leading-order spherical harmonic coefficients in the detector frame for the self-correlation channels with normalization factor $3\sqrt\pi/5$ as
\begin{equation}
\bar b^{ZZ}_{00} = 1, \quad
\bar b^{ZZ}_{20} = \frac{2\sqrt{5}}{7}, \quad
\bar b^{ZZ}_{40} = \frac{1}{42}, \quad
\bar b^{ZZ}_{4\pm 4} = -\frac{\sqrt{70}}{84},
\label{eq:seeds}
\end{equation}
which are related to the other self-correlation channels coefficients by the cyclic relation phase shifts given in Eq.~\eqref{eq:cyclic}, or explicitly
\begin{equation}
    \bar{b}^{ZZ}_{\ell m} = e^{-2 \pi i m/3} \bar{b}^{YY}_{\ell m} = e^{2 \pi i m/3} \bar{b}^{XX}_{\ell m}
\label{eq:explicit-cyclic}
\end{equation}

Substituting $\Ddy_k = -(\Ddy_i+\Ddy_j)$ for equal arms and up to $\mathcal{O}(\chi^2)$ from Eq.~\eqref{eq:detector-tensor-resid} into Eq.~\eqref{eq:Blm{Di,Dj]}} we get
\begin{equation}
\bar b^{\,kk}_{\ell m} = \bar b^{\,ii}_{\ell m} + \bar b^{\,jj}_{\ell m}
+ \bar b^{\,ij}_{\ell m} + \bar b^{\,ji}_{\ell m},
\label{eq:foursum}
\end{equation}
and the cyclic relations from Eq.~\eqref{eq:cyclic} reduce this to
\begin{equation}
\bar b^{\,ij}_{\ell m}+\bar b^{\,ji}_{\ell m}
= \left(1-1/\omega-1/\omega^2\right)\bar b^{\,kk}_{\ell m},
\label{eq:symid}
\end{equation}
which constrains the symmetric part of $\bar b^{\,ij}$. For even-$\ell$, even-$m$ modes the detector tensors are real, $\bar b^{\,ji} = \bar b^{\,ij}$, and Eq.~\eqref{eq:symid} halves. Thus,
\begin{equation}
\bar b^{\,ij}_{\ell m} =
\begin{cases}
-\tfrac{1}{2}\,\bar b^{\,kk}_{\ell m}, & m \equiv 0 \!\!\pmod 3,\\[4pt]
\phantom{-}\bar b^{\,kk}_{\ell m}, & \text{otherwise,}
\end{cases}
\qquad k \neq i,j .
\label{eq:crossself}
\end{equation}
Thus, the leading-order response in the detector frame only has support for the $\ell = 0, 2, \text{and } 4$ modes at $|m| = 0, 4$. The Wigner rotations described in Sec. \ref{subsec:Wigner} rotate the coefficients to the barycentric frame, smear the response power over all $m \in [-\ell, \ell]$ values, but preserve the $\ell$ modes. Thus, the leading-order response in the barycentric frame has support for $\ell = 0, 2, \text{and } 4$ modes with $m \in [-\ell, \ell]$.

\subsubsection{Signal orthogonality of the $A$, $E$, $T$ basis} \label{subsubsec:aet}
\begin{figure}[b]
\centering
\includegraphics[width=\columnwidth]{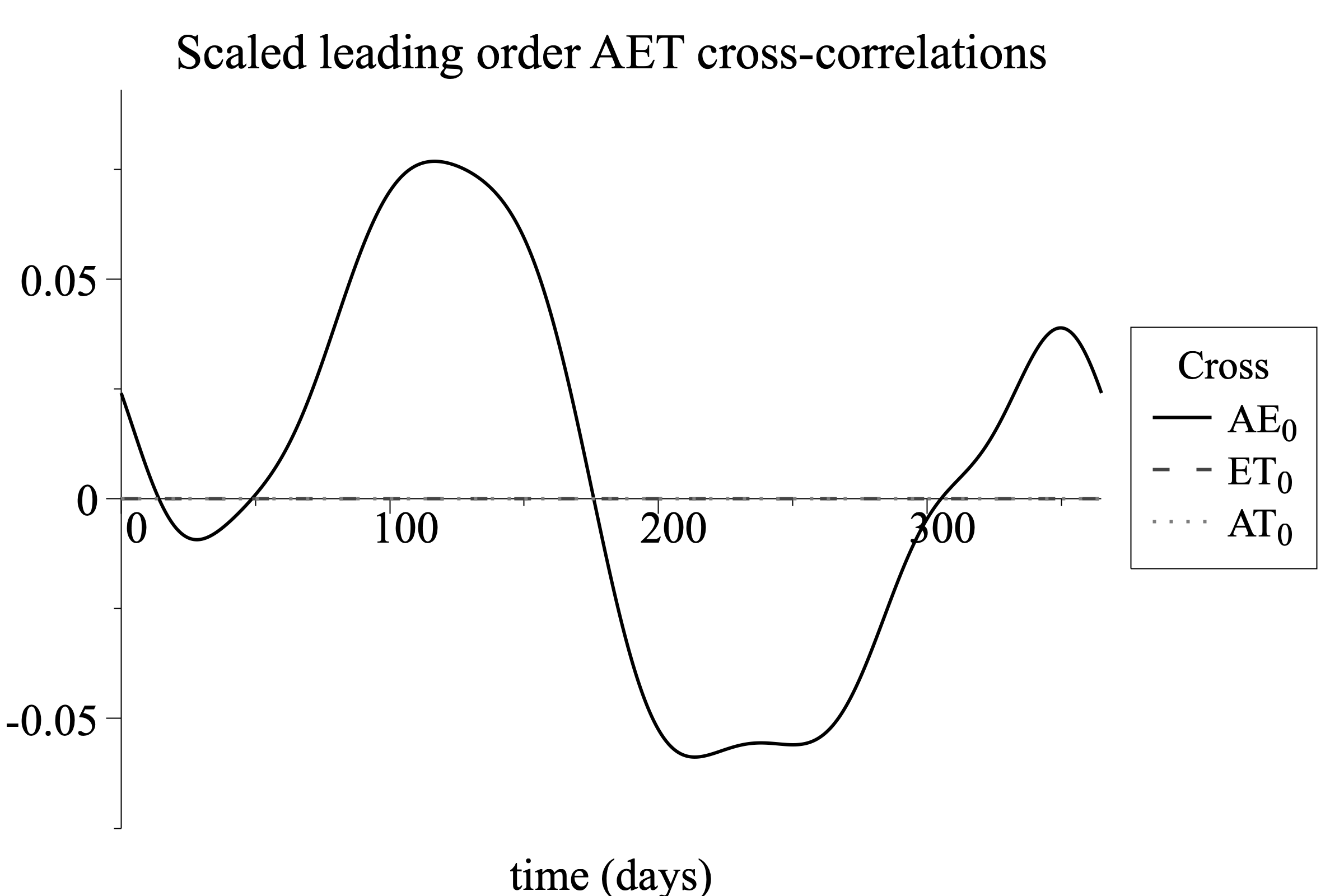}
\caption{Scaled leading-order modulation of the Galactic foreground in the $AE, ET, AT$ cross-correlation channels for the fiducial Galaxy model described later in Sec. \ref{subsec:healpix}. The modulation is scaled by year and channel-averaged normalization term $ \langle \tfrac{1}{3}\langle AA + EE + TT\rangle\rangle_t$. The $AT$ and $ET$ curves vanish identically as the $T$ detector tensor is null in the equal-arm limit. However, $AE$ does not vanish as its detector frame seed is non-zero for the $(4, \pm4)$ mode. The $AE$ response then measures the $\ell = 4$ power of the Galaxy and does not vanish for an anisotropic sky.}
\label{fig:aet}
\end{figure}
The $A$, $E$, $T$ channel combinations are constructed to diagonalize the instrument noise covariance \cite{Hartwig2023}. However, they do not diagonalize the signal covariance for the leading-order, equal-arm response. Thus, we work with the full $3 \times 3$ complex covariance matrix in the $X,Y,Z$ basis.

The $A, E, T$ channels can be written in terms of the $X, Y, Z$ channels \cite{Prince2002, Braglia_2024} in the equal-arm limit as 
\begin{equation}
A=\frac{Z-X}{\sqrt{2}},
\quad
E=\frac{X-2\,Y + Z}{\sqrt{6}},
\quad
T=\frac{X+Y+Z}{\sqrt{3}}.
\label{eq:aetdef}
\end{equation}
and their corresponding detector tensors follow from linear combinations of $\Ddy_X, \Ddy_Y, \Ddy_Z$. In the equal-arm limit, and up to $\mathcal{O}(\chi^2)$, Eq.~\eqref{eq:detector-tensor-resid} gives $\Ddy_X + \Ddy_Y + \Ddy_Z = 0$,
\begin{equation}
\Ddy_T = 0, \qquad \Ddy_E = -\sqrt{\tfrac32}\,\Ddy_Y .
\label{eq:aet-tensors}
\end{equation}
Consequently $\bar b^{AT}_{\ell m} = \bar b^{ET}_{\ell m} = \bar b^{TT}_{\ell m} = 0$, and the only surviving off-diagonal corresponds to the $AE$ channel. Using Eq.~\eqref{eq:aet-tensors} and the sesquilinearity of $\mathcal{B}_{\ell m}[\boldsymbol U, \boldsymbol V]$,
\begin{equation}
\bar b^{AE}_{\ell m}
= \mathcal{N}\,\mathcal{B}_{\ell m}\!\left[\Ddy_A, \Ddy_E\right]
= \frac{\sqrt3}{2}\left(\bar b^{XY}_{\ell m} - \bar b^{ZY}_{\ell m}\right).
\label{eq:bAE}
\end{equation}
As the third index is $Z$ for $XY$ and $X$ for $ZY$, Eq.~\eqref{eq:crossself} carries both terms to self-correlation coefficients, and Eq.~\eqref{eq:explicit-cyclic} gives their phase shift. Thus,
\begin{equation}
\begin{aligned}
\bar b^{AE}_{\ell m}
& = \frac{\sqrt3}{2}\,c_m\left(1 - e^{-2\pi i m/3}\right)\bar b^{ZZ}_{\ell m}, \\
& \\
&
 c_m =
\begin{cases}
-\tfrac12, & m \equiv 0 \!\!\pmod 3,\\[4pt]
\phantom{-}1, & \text{otherwise.}
\end{cases}
\label{eq:bAE-reduced}
\end{aligned}
\end{equation}
such that the factor $1 - e^{-2\pi i m/3}$ vanishes precisely when $m \equiv 0 \pmod 3$. The leading-order response has support at $|m| = 0, 4$, and $|m| = 4$ occurs solely at $\ell = 4$. Thus, the only non-zero $AE$ leading-order coefficient is
\begin{equation}
 \bar b^{AE}_{4,\pm4} = - \frac{\sqrt{70}}{112}\left(\sqrt3 \pm i\right).
\label{eq:bAE-result}
\end{equation}

The same conclusion follows pointwise on the sky. At leading-order $A$ and $E$ are equivalent to a pair of co-located $90^\circ$ Michelson detectors rotated by $45^\circ$, for which $\Ddy_A = \hat x \hat x - \hat y \hat y$ and $\Ddy_E = \hat x \hat y + \hat y \hat x$. Contracting with the polarization tensors, we get
\begin{equation}
\begin{aligned}
\Ddy_A : \textbf{e}^{+} &= \left(1 + \cos^2\bar\theta\right)\cos 2\bar\phi, &
\Ddy_A : \textbf{e}^{\times} &= -2\cos\bar\theta \sin 2\bar\phi, \\
\Ddy_E : \textbf{e}^{+} &= \left(1 + \cos^2\bar\theta\right)\sin 2\bar\phi, &
\Ddy_E : \textbf{e}^{\times} &= \phantom{-}2\cos\bar\theta \cos 2\bar\phi,
\end{aligned}
\label{eq:aet-patterns}
\end{equation}
and since
$\left(1 + \cos^2\bar\theta\right)^2 - 4\cos^2\bar\theta = \left(1 - \cos^2\bar\theta\right)^2$,
the polarization sum is
\begin{equation}
\sum_A \left(\Ddy_A : \textbf{e}^A\right)\left(\Ddy_E : \textbf{e}^A\right)
= \frac{1}{2}\sin^4\bar\theta \, \sin 4\bar\phi ,
\label{eq:aet-overlap}
\end{equation}
which is a pure $\ell = 4$, $m = \pm4$ angular pattern as $Y_{4,\pm4} \propto \sin^4\bar\theta\, e^{\pm 4 i \bar\phi}$ and orthonormality removes every other $(\ell, m)$ from the integral defining
$\mathcal{B}_{\ell m}[\Ddy_A, \Ddy_E]$.

The $A$ and $E$ responses therefore overlap at every point on the sky and the overlap integrates to zero only against an isotropic weight. A foreground carrying power at $\ell = 4$, as the anisotropic Galaxy does, populates $\langle AE^{*}\rangle$ even for a perfectly equilateral constellation, and the signal covariance is not diagonal in the $A$, $E$, $T$ basis. Fig.~\ref{fig:aet} shows the three leading-order $A, E, T$ basis cross-correlations for the fiducial Galaxy model described later in Sec. \ref{subsec:healpix}.

\subsection{Finite-arm corrections to $\mathcal{O}(f/f_*)^2$}\label{SubSec:FiniteArm}
\label{sec:finite}
We Taylor expand the round-trip transfer function in Eq.~\eqref{eq:Tr} and $\Theta$ with respect to $\chi = f/f_\ast \ll 1$ in the equal-arm limit $\epsilon \rightarrow 0$, to get 
\begin{equation}
\begin{aligned}
& \Tr(\mu) = 1 - \frac{\ii\, \chi}{2}(2+\mu)
- \frac{\chi^2}{6}\left(4 + 3\mu + \mu^2\right) + \mathcal{O}(\chi^3), \\
& \Theta = \ee^{-2\pi \ii f \kh \cdot \xGC}  \left( 1 - \ii \, \chi A_i - \frac{\chi^2 A_i^2}{2} + \mathcal{O}(\chi^3) \right)
\end{aligned}
\label{eq:Trexpand}
\end{equation}
and substitute them in Eq.~\eqref{eq:anntenna} to get the finite armlength antenna pattern $F^A_i(f;\kh)$ and response $R^{jk}$. We then compute spherical harmonic coefficients of the finite-arm corrected response up to $\mathcal{O}(\chi^2)$ with normalization factor $3\sqrt\pi/5$ as
\begin{equation}
\begin{aligned}
& \bar b^{ZZ}_{00} = 1 -\frac{169}{504}\chi^2, 
\quad
\bar b^{ZZ}_{20} = \frac{2\sqrt{5}}{7} - \frac{13 \sqrt{5}}{126} \chi^2, \\
\quad
& \bar b^{ZZ}_{40} = \frac{1}{42} + \frac{61}{11088} \chi^2, \quad
\bar b^{ZZ}_{4\pm 4} = -\frac{\sqrt{70}}{84} + \frac{3 \sqrt{70}}{616} \chi^2,
\end{aligned}
\label{eq:finite-seeds1}
\end{equation}
and support for purely $\mathcal{O}(\chi^2)$ modes at
\begin{equation}
\begin{aligned}
& \bar b^{ZZ}_{2\pm2} = \frac{5 \sqrt{30}}{1512}\chi^2, 
\quad
\bar b^{ZZ}_{4 \pm 2} = - \frac{13 \sqrt{10}}{11088} \chi^2, \\
\quad
& \bar b^{ZZ}_{60} = \frac{5 \sqrt{13}}{48048} \chi^2, \quad \bar b^{ZZ}_{6\pm2} = \frac{\sqrt{1365}}{432432} \chi^2,\quad \\
& \bar b^{ZZ}_{6\pm4} = -\frac{5\sqrt{182}}{72072} \chi^2.
\end{aligned}
\label{eq:finite-seeds2}
\end{equation}
Our finite-arm self-correlation results are consistent with previous studies \cite{Cornish2002, Cornish2001}, which place the two arms at azimuths $5\pi/6$ and $7\pi/6$, an offset of $\gamma_n = 5\pi/6 - 2\pi n/3$ from the arms of our channel $n$ at $\psi = 2\pi n/3$ and $2\pi n/3 + \pi/3$. Rotating our detector-frame coefficients through $\gamma_n$ about $\bar z$ using Eq.~\eqref{eq:cyclicgen} reproduces Eq. (18) of \cite{Cornish2002}. 

At second order in $f/f_\ast$ the angular structure of the response extends to higher multipoles. All the $\ell = 6$ modes and $|m| = 2$ structure across $\ell = 2,4,6$ are entirely $\mathcal{O}(\chi^2)$ effects. Thus, after the Wigner rotation, finite-arm corrected response has support for $\ell_\text{max} = 6$ modes with $m \in [-\ell,\ell]$ where the leading-order response reached only the $\ell_\text{max} = 4$ modes. 

The finite-arm cross-correlation channel coefficients up to $\mathcal{O}(\chi^2)$ for these even-$\ell$, even-$m$ modes can then be calculated using Eq.~\eqref{eq:crossself}. The cross-correlation response also produces purely imaginary odd-$\ell$, odd-m $\mathcal{O}(\chi)$ modes at $\ell=3,5$ and $|m| =3$ for which $\bar b^{\,ii}_{\ell m} = 0$ and $\bar b^{\,ji} = -\bar b^{\,ij}$. Eq.~\eqref{eq:symid} is then uninformative and the anti-symmetric, purely imaginary $\bar b^{\,ij}$ for $i \neq j$ are calculated explicitly as
\begin{equation}
\bar b^{XY}_{3\pm3} = \mp \; \ii \frac{\sqrt{35}}{72} \chi \qquad
\bar b^{XY}_{5\pm3} = \mp \; \ii \frac{\sqrt{385}}{5544} \chi 
\label{eq:finite-seedsIm}
\end{equation}
which are related to the other cross-correlation channels coefficients by the cyclic relation phase shifts given in Eq.~\eqref{eq:cyclic}, or explicitly 
\begin{equation}
    \bar{b}^{XY}_{\ell m} = e^{-2 \pi i m/3} \bar{b}^{ZX}_{\ell m} = e^{2 \pi i m/3} \bar{b}^{YZ}_{\ell m}
\label{eq:explicit-cyclic-cross}
\end{equation}
While the imaginary $\mathcal{O}(\chi)$ cross channel coefficients do not affect the modulation curves, they should be incorporated into the likelihood calculation, Fisher analysis, and quadrature spectrum.

\subsection{Unequal-arm corrections}\label{subsec:unequal}
We account for corrections in the response due to unequal but constant arms $\ell_{ij}=L(1+\epsilon_{ij})$ that are static over the light-crossing time with $\epsilon_{ij}=\epsilon_{ji}$. The arm lengths of the detector and its opening angles both change as a function of time. Spacecraft 2, however, defines the $\bar x$ axis of the vertex-pinned detector frame in Eq.~\eqref{eq:arms} that we used for the equal-arm case. As the arm lengths change, spacecraft 2 as well as the $\bar x$ axis rotate about $\bar z$ by perturbation $\delta \Psi_r(\epsilon_{ij})$ relative to the rigid, equilateral frame. Pinning spacecraft 2 to the $\bar x$ axis makes the rotation to the barycentric frame $R(\alpha,\lambda)\, R_{z}(\delta\Psi_r)$ and the $\alpha$-dependence of the Wigner rotation in Sec. \ref{subsec:Wigner} is no longer a simple exponential giving the selection rule $k=m-n$. 

\begin{figure}
\centering
\begin{tikzpicture}[
    line cap=round,
    >={Stealth[length=2mm]},
    arm/.style   = {thick},
    axis/.style  = {thick},
    ang/.style   = {thin},
    scale=0.85   % fits one APS column; use figure* with scale=1.25 for full width
  ]

  % ---- geometry -----------------------------------------------------------
  % dPsi_r and the arm lengths are exaggerated for legibility.
  \def\dpsi{6}        % azimuth of arm 12 = delta Psi_r  [deg]
  \def\Aang{42}       % azimuth of arm 13              [deg]
  \def\Ltwo{6.4}      % \ell_12
  \def\Lthree{6.2}    % \ell_13

  \coordinate (O)  at (0,0);                         % spacecraft 1
  \coordinate (S2) at (\dpsi:\Ltwo);                 % spacecraft 2
  \coordinate (S3) at (\Aang:\Lthree);               % spacecraft 3

  % ---- axes ---------------------------------------------------------------
  \draw[axis,->] (-0.9,0) -- (7.7,0);
  \draw[axis,->] (0,-1.0) -- (0,4.75);
  \node[below=3pt] at (7.30,0)   {\large$\bar{x}$};   % directly under the axis
  \node[left=3pt]  at (0,4.55)   {\large$\bar{y}$};   % beside the arrowhead

  % shade the (exaggerated) frame-rotation wedge so the sliver is legible
  \fill[black!10] (O) -- (4.2,0) arc (0:\dpsi:4.2) -- cycle;

  % ---- triangle -----------------------------------------------------------
  \draw[arm] (O)  -- (S2);
  \draw[arm] (O)  -- (S3);
  \draw[arm] (S2) -- (S3);

  % ---- interior angles ----------------------------------------------------
  \pic[ang, draw, angle radius=1.45cm]                {angle = S2--O--S3};
  \pic[ang, draw, angle radius=1.05cm]                {angle = S3--S2--O};
  \pic[ang, draw, angle radius=0.95cm]                {angle = O--S3--S2};

  \node at ({(\dpsi+\Aang)/2}:1.95) {$\bar{A}'$};
  \node at ($(S2)+(153:1.45)$)      {$\bar{B}'$};
  \node at ($(S3)+(255:1.35)$)      {$\bar{C}'$};

  % ---- the frame-rotation angle ------------------------------------------
  \draw[ang] (3.9,0) arc (0:\dpsi:3.9);

  % ---- vertex labels ------------------------------------------------------
  \node[below left=1pt]  at (O)  {$1$};
  \node[right=4pt]       at (S2) {$2$};
  \node[above right=1pt] at (S3) {$3$};

  % ---- arm labels ---------------------------------------------------------
  \node at ($(O)!0.5!(S3)  + (-0.42,0.30)$) {$\ell_{13}$};
  \node at ($(O)!0.52!(S2) + (0.00,0.42)$)  {$\ell_{12}$};
  \node at ($(S2)!0.5!(S3) + (0.52,0.12)$)  {$\ell_{23}$};

  % ---- annotation for delta Psi_r ----------------------------------------
  \coordinate (tip) at ({\dpsi/2}:3.9);
  \node (dpsilab) at (4.3,-1.15)
        {$\delta\Psi_{r} = \Psi_{12}$};
  \draw[->, thin] (dpsilab.north west) to[out=120,in=-80] ($(tip)+(0,-0.06)$);
  \node[below] at (dpsilab.south) {\footnotesize azimuthal angle};

  % --- bounding-box padding: breathing room above, tight below ----------
  \path (-1.25,-1.72) rectangle (7.95,5.55);

\end{tikzpicture}
\vspace{-0.45cm}
\caption{Detector-frame geometry for unequal arms. Spacecraft~1 sits at the origin and the interior angles $\bar{A}',\bar{B}',\bar{C}'$ at spacecraft 1, 2 and 3 face the opposite arms $\ell_{23}, \ell_{13}, \ell_{12}$ respectively, and depart from $\pi/3$ for $\ell_{ij}=\ell(1+\epsilon_{ij})$. In the shape-neutral frame the arm $\ell_{12}$ is no longer pinned to $\bar{x}$ but carries azimuth $\Psi_{12}=\delta\Psi_{r}$. Arm-length differences are exaggerated for clarity.}
\label{fig:unequal-arm-geometry}
\end{figure}
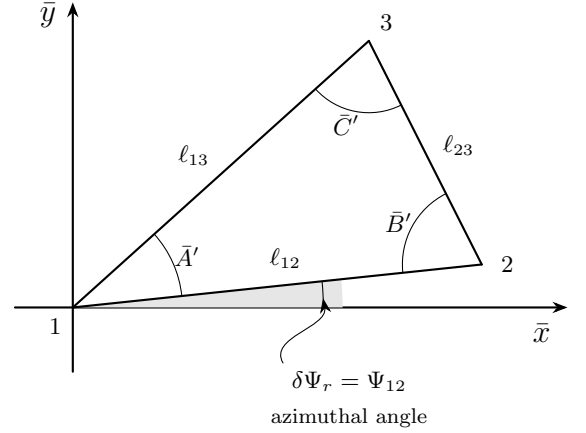

We instead define a shape-neutral frame where spacecraft 2 sits at the azimuthal $\pm \delta \Psi_r$, slightly off the $\bar x$ (Fig.~\ref{fig:unequal-arm-geometry}). This choice preserves the $\mathcal{O}(\epsilon)$ unequal-arm effect in the detector-frame coefficients and keeps the Wigner rotation over a rigid frame. The vertex-pinned and shape-neutral detector-frame coefficients are related by a rotation about $\bar z$ as  $\bar{b}^{\,\mathrm{pin}}_{\ell m} = e^{\,i m \delta\Psi_{r}}\,\bar{b}^{\,\mathrm{SN}}_{\ell m}$. As $\delta\Psi_{r} =\mathcal{O}(\epsilon)$ and the two frames share the same leading-order equal-arm seed $\bar{b}^{(0)}_{\ell m}$, expanding to $\mathcal{O}(\epsilon)$ gives
\begin{equation}
  \delta\bar{b}^{\,\mathrm{pin}}_{\ell m}
  - \delta\bar{b}^{\,\mathrm{SN}}_{\ell m}
  = i m \,\delta\Psi_{r}\, \bar{b}^{(0)}_{\ell m}.
  \label{eq:frame-law}
\end{equation}
Thus, the two conventions only differ for the $(4,\pm4)$ detector frame modes.

\subsubsection{Detector geometry in the shape-neutral frame}
The unit arm vectors of the detector in the shape-neutral frame are $\nh_{ij} = (\cos{\Psi_{ij}}, \sin{\Psi_{ij}}, 0)$, where 
\begin{equation}
\begin{aligned}
& \boldsymbol{\Psi} =
\bigl(\underbrace{\delta\Psi_r}_{\Psi_{12}},\
      \underbrace{\bar{A}' + \delta\Psi_r}_{\Psi_{13}},\
      \underbrace{\pi - \bar{B}' + \delta\Psi_r}_{\Psi_{23}}\bigr) \\
& \boldsymbol{\delta\Psi} =
\bigl(\underbrace{\delta\Psi_r}_{\delta\Psi_{12}},\
      \underbrace{\delta\bar{A} + \delta\Psi_r}_{\delta\Psi_{13}},\
      \underbrace{\delta\Psi_r - \delta\bar{B}}_{\delta\Psi_{23}}\bigr),
\end{aligned}
\end{equation}
$\overline{V}_i' \in [\bar{A}', \bar{B}', \bar{C}']$ are the interior angles of vertices 1, 2, and 3 respectively, and $\overline{V_i}' = \pi/3 + \delta \overline{V_i}$ as the triangle is no longer equilateral. We use the law of cosines to write
\begin{equation}
\cos \overline{V_i}' = \frac{\ell_{ij}^2+\ell_{ik}^2-\ell_{jk}^2}{2\,\ell_{ij}\ell_{ik}}
\simeq \frac12\left[1+\epsilon_{ij}+\epsilon_{ik}-2\epsilon_{jk}\right],
\end{equation}
and using $\overline{V}_i = \pi/3$, we get
\begin{equation}
\delta \overline{V}_i = (2\epsilon_{jk}-\epsilon_{ij}-\epsilon_{ik})/\sqrt{3},
\end{equation}
with $\sum_i \delta \overline{V_i} = 0.$ In the vertex-pinned frame the arm azimuths shift by $(0, \delta \bar{A}, - \delta \bar{B})$, giving the mean azimuthal rotation $(\delta \bar{A} - \delta \bar{B})/3$. The shape-neutral frame removes this by rotating every arm by $\delta\Psi_r = \bigl(\delta\bar{B} - \delta\bar{A}\bigr)/3$ such that $(\mathbf{1}\cdot\delta\boldsymbol{\Psi}) = 0$. This gives
\begin{equation}
\delta\Psi_r = (\epsilon_{13}-\epsilon_{23})/\sqrt{3}.
\end{equation}
Thus, unit arm vectors $\nh_{ij}$ carry the changing geometry of the opening angles at $\mathcal{O}(\epsilon)$ and rotate the arm dyads.

\subsubsection{Rotation of the arm dyads}
% ---------------------------------------------------------------------------
% Unequal-arm detector geometry  (recreates the hand-drawn figure, p.6)
%
% Requires in the preamble of main.tex:
%   \usepackage{tikz}
%   \usetikzlibrary{angles,quotes,calc,arrows.meta}
%
% Insert with:  \input{fig_unequal_arm_geometry}
% ---------------------------------------------------------------------------
We can write the unequal-arm perturbation as $\nh_{ij} = \nh^{(0)}_{ij} + \delta\nh_{ij}$, where $\nh^{(0)}_{ij}$ are the equal-arm unit vectors and $\dd \nh_{ij}/\dd \Psi_{ij} = (-\sin{\Psi_{ij}}, \cos{\Psi_{ij}}, 0) = \nh^\perp_{ij}$. Thus, $\delta\nh_{ij} = \delta \Psi_{ij} \nh^\perp_{ij}$. Perturbing $\nh_{ij}$ in $\Pdy_{ij} \equiv \nh_{ij}\otimes\nh_{ij}$, we get
\begin{equation}
    \Pdy_{ij} = \Pdy^{(0)}_{ij} + \delta\Pdy_{ij} + \mathcal{O}(\epsilon^2)
\end{equation}
where
\begin{equation}
\begin{aligned}
& \delta\Pdy_{ij} \equiv \delta\nh_{ij}\otimes\nh_{ij}+\nh_{ij}\otimes\delta\nh_{ij} =\delta\Psi_{ij}\;\Sdy(\Psi_{ij}) \\
& \Sdy(\Psi_{ij}) \equiv \nh_{ij}^\perp \otimes \nh_{ij}^{(0)} + \nh_{ij}^{(0)} \otimes \nh_{ij}^\perp 
\end{aligned}
\end{equation}
Thus, the unequal-arm corrections $\mathcal{O}(\epsilon)$ enter the detector tensor in Eq.~\eqref{eq:detector-tensor} as arm-product weights $w_{Xi}$ that rescale the response as well as opening-angle perturbations that rotate the arm dyads. The unequal-arm expansion is carried out independently of the frequency expansion, and mixed terms of order $\mathcal{O} (\epsilon \chi)$ and $\mathcal{O} (\epsilon \chi^2)$ are discarded. At $\mathcal{O}(\epsilon)$ we set $\chi \to 0$, so the transfer function $\Tr(\mu_{ij}) \to 1$ in the complex detector tensor Eq.~\eqref{eq:detector-tensor}, and the phase factor that enters the correlation channels as $\ee^{-\ii\chi(A_j - A_i)} \to 1$. The $\epsilon$-dependence of both these terms first enters at $\mathcal{O}(\epsilon\chi)$, through $\chi_{ij} = \chi\,(1+\epsilon_{ij})$ and $A_j - A_i$ in Eq.~\eqref{eq:Adiff} respectively, and is discarded. Expanding the complex detector tensor to first order in $\epsilon$ gives us the unequal-arm correction term
\begin{equation}
\delta\Ddy_i(f;\kh) = \tfrac12\Big[
\left(\epsilon_{ij}+\epsilon_{ik}\right)
 \left(\Pdy_{ij}^{(0)}-\Pdy_{ik}^{(0)}\right)
\nonumber
+\ \delta\Pdy_{ij}-\delta\Pdy_{ik}\Big].
\label{eq:dD}
\end{equation}
The unequal-arm correction, thus, changes both the shape and the magnitude of the leading-order response. We compute the unequal-arm corrections $\delta \bar b^{\,jk}_{\ell m}$ for the leading-order $\bar b^{\,jk}_{\ell m}$ in Eqs.~(\ref{eq:seeds})-(\ref{eq:crossself}) normalized by the same $3 \sqrt{\pi}/5$ factor. For the $m=0$ modes, the self-correlation unequal-arm corrections
\begin{equation}
\delta \bar b_{00} = \frac{4}{3}\,\Sigma_\epsilon, \quad
\delta \bar b_{20} = \frac{8\sqrt{5}}{21}\,\Sigma_\epsilon, \quad
\delta \bar b_{40} = \frac{2}{63}\,\Sigma_\epsilon,
\label{eq:dbself}
\end{equation}
are invariant under cyclic relabeling and phase shifts with $\Sigma_\epsilon \equiv \epsilon_{pq}/2 = \epsilon_{12} + \epsilon_{13}+ \epsilon_{23}$. The cross-correlation correction coefficients are
\begin{equation}
\delta \bar b^{X_i X_j}_{00} = \frac{\Xi_{ij}}{6}, \quad
\delta \bar b^{X_i X_j}_{20} = \frac{\sqrt{5}}{21}\,\Xi_{ij}, \quad
\delta \bar b^{X_i X_j}_{40} = \frac{\Xi_{ij}}{252},
\label{eq:dbcross}
\end{equation}
where $\Xi_{ij} \equiv 9 \epsilon_{ij} - 7 \Sigma_\epsilon$ carries the cyclic relabeling. For the $(4,\pm 4)$ mode, the self and cross channel coefficients can be written as
\begin{equation}
\begin{aligned}
& \delta \bar b^{ZZ}_{44} = -\frac{\sqrt{70}}{252} [4 \Sigma_{\epsilon} + \ii \; 2 \sqrt{3} \,\Delta_3] \\
& \delta \bar b^{XY}_{44} = -\frac{\sqrt{70}}{252} [4 \Sigma_{\epsilon} - \ii \; \sqrt{3} \,\Delta_3]
\end{aligned}
\label{eq:dbm4}
\end{equation}
where $\Delta_i = \epsilon_{ij} - \epsilon_{ik}$ carries the cyclic relabeling, the other self and cross terms can be computed using phase shifts in Eqs.~\eqref{eq:explicit-cyclic} and \eqref{eq:explicit-cyclic-cross} respectively, and the $b_{4,-4}$ modes can be computed using Eq.~\eqref{eq:blm-conjugate}.

The  unequal-arm corrections at $\mathcal{O}(\epsilon)$ match the $l, m$ modes of the leading-order response in the detector frame. They also only have support for the $\ell = 0, 2, \text{and } 4$ modes at $|m| = 0, 4$. The Wigner rotation to the barycentric frame then smears the response power over all $m \in [-\ell, \ell]$ values, but preserves the $\ell$ values. 

\subsubsection{Keplerian orbits}
We compute the unequal-arm corrections for Keplerian spacecraft orbits. To linear order in eccentricity  
\begin{align} \label{ovar}
\epsilon_{12}(t) & = \frac{e}{32} \big[ 15 \sin{(\gamma + \pi/6)}  - \cos{(3 \gamma)} \big] \\
\epsilon_{13}(t) & = - \frac{e}{32} \big[15 \sin{(\gamma - \pi/6)} + \cos{(3\gamma)} \big] \\
\epsilon_{23}(t) & = - \frac{e}{32} \big[ 15 \cos{\gamma} + \cos{(3\gamma)} \big],
\end{align}
where $\gamma = (\alpha-\lambda)$ \cite{Rubbo2004}. We expand the detector-frame coefficients in harmonics of the orbital frequency as
\begin{equation}
\bar b^{ij}_{\ell m}(t) = \sum_{\bar k} \bar{\mathcal{C}}^{ij}_{\ell m,\bar k}\;
e^{2\pi \ii \bar k t/T_{\rm yr}},
\label{eq:FourierHarmonics}
\end{equation}
where $\bar{k}$ are the harmonic indices and $\mathcal{C}^{ij}_{\ell m, \bar{k}}$ are the corresponding Fourier coefficients. With Keplerian spacecraft orbits, the self-correlation coefficient corrections at $m=0$ in Eq.~\eqref{eq:dbself} are proportional to
\begin{equation}
\Sigma_\epsilon = -\frac{3e}{32}\cos{3\gamma},
\end{equation}
and contain only the third annual modulation harmonic $\bar{k}=\pm3$ in the detector frame. The cross-correlation coefficient corrections at $m=0$ modes in Eq.~\eqref{eq:dbcross} are proportional to
\begin{equation}
\Xi_{ij} = \frac{3e}{32}\Big[45\cos\!\left(\gamma+\tfrac{\pi}{3} s_{ij}\right) + 4\cos 3\gamma\Big],
\end{equation}
with $(s_{12}, s_{13}, s_{23}) = (-1, 1, 3)$. Thus, they retain both the first and the third annual modulation harmonic $\bar{k}=\pm1, \pm3$ in the detector frame, with the first harmonic $45/4 = 11.25$ times the amplitude of the third harmonic. The $|m|=4$ modes for the self and cross coefficient corrections from Eq.~\eqref{eq:dbm4} 
\begin{equation}
\begin{aligned}
& \delta \bar b^{ZZ}_{44} = \frac{\sqrt{70} \, e}{672}\left[\cos 3\gamma + \ii \, \frac{15}{2} \sin\!\left(\gamma - \frac{\pi}{3}\right)\right] \\
& \delta \bar b^{XY}_{44} = \frac{\sqrt{70} \, e}{672}\left[\cos 3\gamma - \ii \, \frac{15}{4} \sin\!\left(\gamma - \frac{\pi}{3}\right)\right]
\end{aligned}
\end{equation}
also contain the annual modulation harmonics $\bar{k}=\pm1, \pm3$ in the detector frame. The Wigner rotation to the barycentric frame (Sec. \ref{subsec:Wigner}) acting on these time-dependent coefficients then shifts $k$ by $\bar{k} = \pm1, \pm 3$, extending the response's annual modulation harmonic content from $|k|\le8$ to $|k|\le11$.

\subsection{Restoring the annual modulation}\label{subsec:Wigner}

\subsubsection{Wigner D rotations}
We map the harmonic coefficients in the detector frame $\bar{b}_{\ell m}^{jk}$ to the harmonic coefficients in the barycentric frame $b_{\ell m}^{jk}$ using Wigner D matrices as
\begin{equation}
b_{\ell m} = \sum_{n=-\ell}^\ell \boldsymbol{D}_{mn}^\ell(\varphi, \vartheta, \psi) \; \bar{b}_{\ell n}
\label{eq:WignerD}
\end{equation}
where the Euler angle convention $(\psi, \vartheta, \varphi)$ performs a series of right-handed rotations, starting with a rotation by $\psi$ about the original $z$ axis, followed by rotation by $\vartheta$ about the original $y$ axis, followed by a rotation by $\varphi$ about the original $z$ axis \cite{Varshalovich1988, Cornish2002}. The D matrices are defined as
\begin{equation}
\boldsymbol{D}^{\ell}_{nm}(\psi,\vartheta,\varphi)
= \ee^{-\ii n \psi}\, d^{\ell}_{nm}(\vartheta)\, \ee^{-\ii m\varphi},
\label{eq:Ddef}
\end{equation}
where
\begin{align}
d_{nm}^{\ell}(\vartheta)
&=
\sum_i (-1)^i
\frac{\left[(\ell+n)!(\ell-n)!(\ell+m)!(\ell-m)!\right]^{1/2}}
{(\ell+n-i)!(\ell-m-i)!i!(i+m-n)!}
\nonumber\\
&\qquad\times
\left[\cos(\vartheta/2)\right]^{2\ell-2i+n-m}
\left[\sin(\vartheta/2)\right]^{2i+m-n},
\end{align}
and $\ell,n$ are the spherical harmonic modes in the original frame, or in our case the detector frame \cite{Cornish2002}. Using the identity ${D}_{mn}^\ell(\varphi, \vartheta, \psi) = {D}_{nm}^\ell(\psi, -\vartheta, \varphi)$ \cite{Varshalovich1988}, we get 
\begin{equation}
b_{\ell m} = \sum_{n=-\ell}^\ell \ee^{-\ii n \psi}\, d^{\ell}_{nm}(-\vartheta)\, \ee^{-\ii m\varphi} \; \bar{b}_{\ell n}
\label{eq:wigner}
\end{equation}
which gives the barycentric coefficients directly from detector frame coefficients and the Euler angles.

\subsubsection{Euler angles and orientation of the constellation}

The spacecraft follow the standard eccentric, inclined Keplerian orbits \cite{Rubbo2004}, expanded to first order in the eccentricity $e$,
\begin{align}
x_n &= R\cos\alpha
     + \tfrac{1}{2}eR\big[\cos(2\alpha-\beta_n) - 3\cos\beta_n\big],
     \nonumber\\
y_n &= R\sin\alpha
     + \tfrac{1}{2}eR\big[\sin(2\alpha-\beta_n) - 3\sin\beta_n\big],
     \label{eq:orbits}\\
z_n &= -\sqrt{3}\,eR\cos(\alpha-\beta_n),
     \nonumber
\end{align}
with $\alpha = 2\pi f_m t + \kappa$ the orbital phase of the guiding center, $\beta_n = 2\pi n/3 + \lambda$ the phase of spacecraft $n$ within the constellation for $n = 0,1,2$, and $e = L/(2\sqrt{3}R)$ fixed by the nominal arm length.  The separation vectors between the spacecrafts can then be written as 
\begin{equation}
{\mathbf{x}}_m - {\mathbf{x}}_n
  = 4 eR\sin\delta\,\big(\cos u\,\mathbf{\hat{e}}_1 + \sin u\,\mathbf{\hat{e}}_2\big),
\label{eq:armdecomp}
\end{equation}
where $\sigma \equiv (\beta_m+\beta_n)/2$, $\delta \equiv (\beta_m-\beta_n)/2$
and $u \equiv \alpha-\sigma$ and the basis vectors in the constellation plane
\begin{equation}
\mathbf{\hat{e}}_1 = \begin{pmatrix}\sin\alpha\\ -\cos\alpha\\ 0\end{pmatrix},
\qquad
\mathbf{\hat{e}}_2 = - \frac{1}{2}\begin{pmatrix}\cos\alpha\\ \sin\alpha\\ \sqrt{3}\end{pmatrix}.
\label{eq:e1e2}
\end{equation}
We can then express the constellation frame $\bar x$, $\bar y$, $\bar z$ in the barycentric frame as
\begin{align}
\hat{\bar z} &= -\hat{\mathbf{n}}  \equiv - \hat{\mathbf{e}}_1\times\hat{\mathbf{e}}_2
   = \big(-\tfrac{\sqrt3}{2}\cos\alpha,\,
-\tfrac{\sqrt3}{2}\sin\alpha,\, \tfrac12\big),
  \label{eq:zbar}\\[3pt]
\hat{\bar x} &= \frac{\mathbf{x}_1-\mathbf{x}_0}{|\mathbf{x}_1-\mathbf{x}_0|}
   = \cos u_0\,\hat{\mathbf{e}}_1 + \sin u_0\,\hat{\mathbf{e}}_2,
  \label{eq:xbar}\\[3pt]
\hat{\bar y} &= \hat{\bar z} \times \hat{\bar x}
   = \sin u_0\,\hat{\mathbf{e}}_1 - \cos u_0\,\hat{\mathbf{e}}_2 ,
  \label{eq:ybar}
\end{align}
where the unit vectors are independent of the choice of origin and $u_0 = \alpha - \lambda - \pi/3$. As $\hat{\mathbf{e}}_1\cdot\hat z = 0$ and $\hat{\mathbf{e}}_2\cdot\hat z = -\tfrac{\sqrt3}{2}$, only the $\hat{\mathbf{e}}_2$ components survive projection onto $\hat z$:
\begin{equation}
\begin{aligned}
\hat{\bar x}\cdot\hat z
  &= -\tfrac{\sqrt3}{2}\sin u_0
   = \tfrac{\sqrt{3}}{2}\sin\!\big(\lambda-\alpha+\tfrac{\pi}{3}\big),\\
\hat{\bar y}\cdot\hat z
  &= \tfrac{\sqrt3}{2}\cos u_0
   = \tfrac{\sqrt{3}}{2}\cos\!\big(\lambda-\alpha+\tfrac{\pi}{3}\big),\\
\hat{\bar z}\cdot\hat z &= \tfrac12 .
\end{aligned}
\label{eq:thirdrow}
\end{equation}
We adopt the Euler convention of right-handed rotations $\mathbb{R} = R_z(\varphi)R_y(\vartheta)R_z(\psi)$ \cite{Cornish2002} which carry the constellation frame  components to the barycentric frame. Abbreviating $c_x \equiv \cos x$ and $s_x \equiv \sin x$,
\begin{equation}
\setlength{\arraycolsep}{2.5pt}
\mathbb{R} =
\begin{pmatrix}
c_\varphi c_\psi c_\vartheta - s_\varphi s_\psi &
-c_\varphi s_\psi c_\vartheta - s_\varphi c_\psi &
c_\varphi s_\vartheta \\[3pt]
c_\psi c_\vartheta s_\varphi + c_\varphi s_\psi &
-s_\varphi s_\psi c_\vartheta + c_\varphi c_\psi &
s_\varphi s_\vartheta \\[3pt]
-s_\vartheta c_\psi &
s_\vartheta s_\psi &
c_\vartheta
\end{pmatrix},
\label{eq:Rgeneral}
\end{equation}
where $\mathbf{x} = \mathbb{R} \cdot \overline{\mathbf{x}}$. The Euler angles are then derived from the rotation matrix and
Eq.~\eqref{eq:thirdrow} as
\begin{align}
c_\vartheta &= \mathbb{R}_{33} = \hat z \cdot \bar z = \tfrac12,
  \nonumber\\[2pt]
s_\vartheta c_\varphi &= \mathbb{R}_{13} = \hat x\cdot\bar z
   = -\tfrac{\sqrt3}{2}\cos\alpha,
  \nonumber\\[2pt]
s_\vartheta s_\varphi &= \mathbb{R}_{23} = \hat y\cdot\bar z
   = -\tfrac{\sqrt3}{2}\sin\alpha,
  \label{eq:match}\\[2pt]
-s_\vartheta c_\psi &= \mathbb{R}_{31} = \hat z \cdot \bar x
   = \tfrac{\sqrt3}{2}\sin\!\big(\lambda-\alpha+\tfrac{\pi}{3}\big),
  \nonumber\\[2pt]
s_\vartheta s_\psi &= \mathbb{R}_{32} = \hat z \cdot \bar y
   = \tfrac{\sqrt3}{2}\cos\!\big(\lambda-\alpha+\tfrac{\pi}{3}\big),
  \nonumber
\end{align}
which gives
\begin{align}
\vartheta &= \arccos(1/2) = \frac{\pi}{3},
  \nonumber\\[2pt]
\varphi   &= \operatorname{atan2}\big(\mathbb{R}_{23},\,\mathbb{R}_{13}\big)
           = \alpha+\pi,
  \label{eq:euler}\\[2pt]
\psi      &= \operatorname{atan2}\big(\mathbb{R}_{32},\,-\mathbb{R}_{31}\big)
           = \lambda - \alpha + \frac{5\pi}{6}.
  \nonumber
\end{align}

The Euler angle \(\vartheta\) is the inclination of the constellation plane relative to the ecliptic. The angle \(\varphi\) specifies the ecliptic longitude of the constellation normal and precesses once per year. Finally, \(\psi\) specifies the rotation of the triangular constellation within its own plane, where the \(-\alpha\) term describes the familiar retrograde cartwheel of the constellation, with one revolution per year opposite to the precession of the normal, \(\lambda\) fixes the initial phase of the triangle, while the constant offset \(5\pi/6\) follows from our chosen convention \(\bar x\parallel\mathbf{x}_1-\mathbf{x}_0\).

Ref.~\cite{Cornish2002} evaluates the self-correlated response of a single Michelson channel, with both arm directions defined relative to a common vertex. The calculation therefore does not incorporate the relative orientation of distinct channels or the initial phase parameter \(\lambda\). In contrast, we retain the cross-channel response and explicitly track the relative spacecraft orientations through \(\lambda\). The two treatments are nevertheless consistent as the reparameterization $ (\varphi,\vartheta,\psi)\rightarrow(\varphi+\pi,-\vartheta,\psi+\pi) $ leaves \(\mathbb R\) unchanged, and Eq.~\eqref{eq:euler} reproduces the Euler angles of Ref.~\cite{Cornish2002} for the implicit initial constellation orientation \(\lambda=\pi/6\).

\subsubsection{Rotating to the barycentric frame}
The detector-frame coefficients therefore rotate as
\begin{equation}
b^{jk}_{\ell m}(\alpha,\lambda)
  = \sum_{n=-\ell}^{\ell}
    D^{\ell}_{nm}\!\left(\psi,\,-\tfrac{\pi}{3},\,\alpha+\pi\right)
    \bar b^{jk}_{\ell n},
\label{eq:brot}
\end{equation}
and with Euler angles $\psi = 5 \pi/6 +\lambda-\alpha, \vartheta = \pi/3, \varphi = \alpha + \pi$, we get
\begin{align}
b_{\ell m}(\alpha,\lambda)
&= \ee^{-\ii m (\alpha + \pi)}
\sum_{n=-\ell}^{\ell}
d^{\ell}_{nm}\!\left( -\tfrac{\pi}{3}\right)
\ee^{-\ii n (5 \pi/6 + \lambda -\alpha)}\,
\bar b_{\ell n}
\label{eq:wignerapp}
\end{align}
Here $\lambda \in \{0, 2\pi/3,
4\pi/3\}$ selects which of the three Michelson vertices is referenced.

Rotation preserves $\ell$ but spreads power across $|m|\le\ell$. The barycentric coefficients only depend on $\alpha$ through $\ee^{-\ii m\alpha}\ee^{+\ii n\alpha}$, constraining the annual harmonic index
\begin{equation}
k = m - n
\label{eq:wignerk}
\end{equation}
where $|k| \le 2\ell$ for time-independent seeds, such as those of the leading-order and finite armlength correction coefficients. 

\subsubsection{Worked example: how $m=\pm1$ appears}
\label{sec:l2example}
In the equal-arm limit, the $\ell = 2$ mode in the detector frame $\bar{b}_{2n}$ has support at $n = 0$ only at leading-order and acquires $n = \pm2$ at $\mathcal{O}(\chi^2)$ as shown in Eq.~\eqref{eq:finite-seeds2}. It does not have support at $n = \pm1$. Yet, the barycentric coefficients $b_{2,\pm1}(\alpha,\lambda)$ are non-zero at both the leading and $\mathcal{O}(\chi^2)$ orders and are generated entirely by the Wigner rotation.

Writing Eq.~\eqref{eq:wigner} for $\ell = 2$, $m = 1$ in terms of the non-zero detector frame seeds, we get 
\begin{equation}
b_{21}
= \ee^{-\ii\varphi}\Big[\,
 d^{2}_{01}\,\bar b_{20}
+d^{2}_{21}\,\ee^{-2\ii\psi}\,\bar b_{22} +d^{2}_{-2,1}\,\ee^{+2\ii\psi}\,\bar b_{2,-2}\Big],
\label{eq:b21}
\end{equation}
where the $d_{nm}^{\ell}(-\pi/3)$ are given as
\begin{equation}
d^{2}_{01} = -\frac{3\sqrt2}{8},\qquad
d^{2}_{21} = \frac{3\sqrt3}{8},\qquad
d^{2}_{-2,1} = -\frac{\sqrt3}{8},
\end{equation}
and the identity $\bar b_{\ell,-n} = (-1)^n \bar
b^{\,*}_{\ell n}$ gives $\bar b_{2,-2} = \bar b^{\,*}_{22}$. Thus,
\begin{equation}
b_{21}=
-\frac{e^{-i\varphi}}{8}
\left[
3\sqrt{2}\,\bar b_{20}
-3\sqrt{3}\,e^{-2i\psi}\bar b_{22}
+\sqrt{3}\,e^{2i\psi}\bar b_{22}^{*}
\right].
\label{eq:b21final}
\end{equation}
and $b_{2,-1} = -b^{\,*}_{21}$.
Note that the $\bar b_{20}$ term survives when $\bar b_{22} = 0$. A purely axisymmetric seed still produces $m = \pm1$ in the barycentric frame, because the rotation axis is not the seed's symmetry axis. The absence of an $n$ mode in the detector frame says nothing about its presence in the rotated frame. 

\subsection{Parameterized Galaxy to foreground multipoles}
\label{subsec:healpix}

We adopt a notional two-component disk-and-bulge model with luminosity density
\begin{equation}
\rho(u,z) = A\,\ee^{-r^2/R_b^2}
          + (1-A)\,\ee^{-u/R_d}\,\mathrm{sech}^2\!\left(z/Z_d\right),
\label{eq:density}
\end{equation}
described by five parameters, $\{A,R_b,R_d,Z_d,R_{\rm GC}\}$ where $A$ is the dimensionless bulge fraction confined to $[0,1]$, $R_b$ is the bulge scale radius, $R_d$ is the disk scale width, $Z_d$ is the disk scale height, $R_{GC}$ is distance to the Galactic center, and $r = \sqrt{x^2 + y^2 + z^2}$ and $u = \sqrt{x^2 + y^2}$ are the Galactocentric spherical and cylindrical radii \cite{Adams2014}. Our fiducial values are $A = 0.25$, $R_b = 0.8$~kpc, $R_d = 2.5$~kpc, $Z_d = 0.6$~kpc and $R_{\rm GC} = 8.2$~kpc. We model the population as equal-luminosity emitters, where the flux received from a volume element is $\rho\,\dd V/\mathscr{r}^2 = \rho\,\dd \mathscr{r}\,\dd\Omega$. Note that $\mathcal{L}$ is invariant, up to a normalization factor, under a common rescaling of $R_b$, $R_d$, $Z_d$ and $R_{\rm GC}$. The modulation, therefore, only constrains ratios of lengths. We do not impose a heliocentric exclusion radius, as its effect on the modulation curves is degenerate with that of $R_{\rm GC}$.

$\mathcal{L}$ is evaluated on a HEALPix grid \cite{Gorski2005} at
$\texttt{Nside}=32$ in ecliptic coordinates, where each pixel direction is rotated into the Galactic frame before performing the radial integral. As the HEALPix resolution is increased, the recovered low-\(\ell\) Galaxy harmonic coefficients rapidly converge. Doubling the resolution changes the \(a_{\ell m}\) vector by approximately 50\% between $\texttt{Nside}=4$ and 8, 2.9\% between 8 and 16, and 0.33\% between 16 and 32. Thus, all relevant large-scale Galaxy information converges by $\texttt{Nside}=16$, with $\texttt{Nside}=32$ providing a conservative choice. The radial integral uses an adaptive Simpson rule at a relative tolerance of $10^{-6}$, such that the sampling concentrates across the bulge and the inner disk, where the density varies rapidly, and coarsens in the outer disk. The map is projected onto spherical harmonics through $\ell_{\max}=6$ by direct pixel summation, with two rounds of iterative refinement. 

The detector coefficients $b^{jk}_{\ell m}(t)$ depend only on the constellation and are computed once. Thus, a new Galaxy costs a sky map evaluation, spherical harmonic transform and contraction of Eq.~\eqref{eq:factorization}. The sky map costs $12 \;\texttt{Nside}^2$ line-of-sight integrals, one per pixel, each an adaptive Simpson quadrature taking of order $10^3$ evaluations of the density model. Projecting the map onto spherical harmonics costs $5 N_{\rm pix} N_{\ell m}$ operations, where  $N_{\ell m} = (\ell_{\rm max} + 1)(\ell_{\rm max} + 2)/2 = 28$ at $\ell_{\rm max} = 6$, and the factor of five accounting for the initial quadrature and two refinement iterations. Contracting the $a_{\ell m}$ against the response coefficients then costs $N_\alpha N_b \sum_\ell (2\ell + 1)^2$ where $N_\alpha = 512$ orbital phases span the year, $\sum_\ell (2\ell+1)^2 = 455$ counts the pairs of azimuthal indices coupled by the Wigner rotation at each multipole, and $N_b = 20$ is the number of non-zero $b_{\ell m}$ coefficients, including those for the finite and unequal-arm corrections. In terms of CPU time, using typical 2024 hardware, each line-of-sight integral takes $10.4~\mu$s, the full map $0.128$~s, the harmonic projection $3.1$~ms, and the contraction $2.8$~ms. The modulation curve for a single channel, corrected for finite and unequal arms, takes about $0.134$~s, most of which is spent on the line-of-sight integration through the density model. Subsequent channel evaluations take another $2.8$~ms. This fast, forward modeling of the unresolved foreground modulation curves enables Fisher analysis for the Galactic parameters.

\subsection{Fisher analysis on the Galactic parameters}
\label{subsec:fisher}
% ---------------------------------------------------------------------- 

As the Galactic foreground is cyclostationary rather than stationary, the wavelet domain  provides a natural basis in which the covariance depends on time and frequency separately~\cite{Cornish:2020odn,Cornish:2026tjt}. The data are tiled into pixels of bandwidth $\Delta f$ and duration $\Delta t$ with $\Delta f\,\Delta t = 1/2$. Let $\bm{d}_{ft}=(X,Y,Z)_{ft}$ denote the three Michelson channels in the time-frequency representation at the pixel $(f,t)$. As the Galactic foreground is the superposition of $\mathcal{O}(10^{7})$ unresolved double white dwarf binaries with independent orbital phases, no deterministic template describes the foreground, the mean $\langle\bm{d}_{ft}\rangle=0$, and all Galactic information is contained in the second moment of the data. The wavelet coefficients for the large number of contributing sources are approximately Gaussian by the central limit theorem. 

As the foreground is slow-varying across the time-frequency tiles, the WDM pixels are uncorrelated up to $\mathcal{O}(\Delta f\,\partial_f\ln S,\ \Delta t\,\partial_t\ln S)$~\cite{Cornish:2025awt}. The covariance is then block diagonal, with a $3\times3$ channel covariance $\mathsf{C}_{ft}$ at each pixel. The joint probability density of the data given the parameters $\bm{q}$ factorizes over the pixels 
\begin{equation}
p(\bm{d}\mid\bm{q}) =
\prod_{f,t} p(\bm{d}_{ft}\mid\bm{q}),
\label{eq:factorise}
\end{equation}
and the probability density for each pixel is given by a zero-mean tri-variate Gaussian as
\begin{equation}
p(\bm{d}_{ft}\mid\bm{q}) =
\frac{1}{(2\pi)^{3/2}\lvert\mathsf{C}_{ft}\rvert^{1/2}}
\exp\!\left[
-\frac{1}{2}\bm{d}_{ft}^{\dagger}
\mathsf{C}_{ft}^{-1}\bm{d}_{ft}
\right],
\label{eq:trivariate}
\end{equation}
where the covariance $\mathsf{C}_{ft}(\bm{q})$ depends on the parameters. For the Galactic foreground, as $\bm{\mu} =0$ and the covariance is parameter-dependent, the log likelihood $\mathsf{L}\equiv-\ln p$ is given as
\begin{equation}
\mathsf{L}
=
\frac{1}{2}\sum_{f,t}
\left[
\ln\det\mathsf{C}_{ft}(\bm{q})
+
\bm{d}_{ft}^{\dagger}
\mathsf{C}_{ft}^{-1}(\bm{q})
\bm{d}_{ft}
\right],
\label{eq:likegal}
\end{equation}
where the Galactic parameters enter through the determinant and inverse of the covariance \cite{TegmarkTaylorHeavens1997}.

In contrast, for a resolvable binary, the parameters enter through the mean $\bm{\mu}=\bm{h}(\bm{q})$, while the covariance is parameter independent. The quadratic term reduces to the familiar form
\begin{equation}
(\bm{d}-\bm{h})^{\dagger}\mathsf{C}^{-1}(\bm{d}-\bm{h})
=
\bm{d}^{\dagger}\mathsf{C}^{-1}\bm{d}
-
2\,
\bm{h}^{\dagger} \mathsf{C}^{-1}\bm{d}
+
\bm{h}^{\dagger}\mathsf{C}^{-1}\bm{h},
\label{eq:binaryexp}
\end{equation}
where the terms $\ln\det\mathsf{C}$ and $\bm{d}^{\dagger}\mathsf{C}^{-1}\bm{d}$ are independent of the source parameters and the term linear in data corresponds to the matched filter.

The Fisher matrix is defined from the log-likelihood as 
\begin{equation}
\Gamma_{ab}
\equiv
\left\langle
\frac{\partial^{2}\mathsf{L}}
{\partial q^{a}\partial q^{b}}
\right\rangle
\end{equation}
and gives
\begin{equation}
\Gamma_{ab}
=
\frac{1}{2}\operatorname{Tr}\!\left[
\mathsf{A}_{a}\mathsf{A}_{b}
+
\mathsf{C}^{-1}\mathsf{M}_{ab}
\right],
\label{eq:tth15}
\end{equation}
where
\begin{equation}
\mathsf{A}_{a}
\equiv
\mathsf{C}^{-1}\mathsf{C}_{,a},
\qquad
\mathsf{M}_{ab}
\equiv
\bm{\mu}_{,a}\bm{\mu}_{,b}^{\dagger}
+
\bm{\mu}_{,b}\bm{\mu}_{,a}^{\dagger},
\label{eq:AM}
\end{equation}
are the covariance and mean-dependent terms respectively \cite{TegmarkTaylorHeavens1997}. A resolvable binary has $\mathsf{C}_{,a}=0$, so the covariance contribution vanishes and the mean dependent term reduces to the familiar noise-weighted inner product, $\Gamma_{ab}=(\partial_a\bm{h}\mid\partial_b\bm{h})$. 

For the unresolved Galactic foreground, $\bm{\mu}=0$, thus $\mathsf{M}_{ab}=0$. The Fisher information is then determined entirely by derivatives of the covariance, which gives
\begin{equation}
\Gamma_{ab} = \frac{1}{2}\sum_{f,t} \operatorname{Tr}\!\left[ \mathsf{C}_{ft}^{-1} \frac{\partial\mathsf{C}_{ft}}{\partial q^{a}} \mathsf{C}_{ft}^{-1} \frac{\partial\mathsf{C}_{ft}}{\partial q^{b}} \right].
\label{eq:fisher}
\end{equation}
As the instrument noise and the Galactic foreground are uncorrelated, the covariance can be written as
\begin{equation}
\mathsf{C}(f,t;\bm{q})=\mathsf{C}_{\rm inst}(f) + \mathsf{C}_{\rm gal}(f,t;\bm{q}),
\label{eq:covsplit}
\end{equation}
where Eq.~\eqref{eq:factorization} separates the Galactic term into a spectral normalization and a modulation matrix
\begin{equation}
\mathsf{C}_{\rm gal}(f,t;\bm{q}) = S_{\rm gal}(f)\, \mathcal{M}(f,t;\bm{q})
\label{eq:covmodel}
\end{equation}
and
\begin{equation}
\mathcal{M}_{X_j X_k}(f,t;\bm{q}) = \frac{1}{\mathcal{N}} \sum_{\ell m} a_{\ell m}^{*}(\bm{q})\, b_{\ell m}^{jk}(f,t),
\label{eq:modmatrix}
\end{equation}
contains the foreground's angular structure and annual modulation, while $S_{\rm gal}(f)$ carries the population-averaged source spectrum. We normalize $\mathcal{M}$ at each frequency by its year-averaged self-correlation, 
\begin{equation}
    \mathcal{N} \equiv \langle \tfrac{1}{3}(\mathcal{M}_{XX}+\mathcal{M}_{YY}+\mathcal{M}_{ZZ})\rangle_t,
\label{eq:modmatrixNorm}
\end{equation}
such that $S_{\rm gal}(f)$ carries the absolute scale. As only $\mathcal{M}$ depends on $\bm{q}$, we get $\partial_a\mathsf{C}=S_{\rm gal}\,\partial_a\mathcal{M}$. As $\mathcal{M}$ is normalized and depends only on the angular structure of the foreground, the Fisher matrix is nearly invariant under a common rescaling of all Galactic length scales $(R_b,R_d,Z_d,R_{\rm GC}) \rightarrow \lambda(R_b,R_d,Z_d,R_{\rm GC})$. For equal-luminosity emitters, this rescaling multiplies the line-of-sight luminosity $\mathcal{L}(\theta,\phi)$ by a constant factor without altering its angular distribution, and is absorbed by the normalization $\mathcal{N}$. The unresolved foreground therefore carries no information on absolute distance scales. A compact Galaxy with a nearby center produces the same angular distribution, and hence the same modulation, as a proportionally larger Galaxy at a greater distance. In the logarithmic basis $(A,\ln R_b,\ln R_d,\ln Z_d,\ln R_{\rm GC})$, the least constrained eigenvector is approximately $\tfrac12(0,1,1,1,1)$, corresponding to an equal shift of all length scales at fixed bulge fraction. We therefore project onto the basis $\left(A,\,\ln\!\left(R_b/R_{\rm GC}\right),\, \ln\!\left(R_d/R_{\rm GC}\right),\, \ln\!\left(Z_d/R_{\rm GC}\right) \right)$, which removes the nearly degenerate overall scale direction and isolates the independent combinations constrained by the foreground modulation. Resolved binaries can provide the absolute length scale required to break this degeneracy, as the frequency, frequency derivative, and amplitude of a detached binary determine its distance from the Sun \cite{Adams2012,Korol2019}.

We adopt the spectral fit
\begin{equation}
S_{\rm gal}^{AE}(f)
=
\frac{\mathcal{A} f^{5/3}}{2}
e^{-(f/f_1)^{\alpha}}
\left[
1+\tanh\!\left(
\frac{f_{\rm knee}-f}{f_2}
\right)
\right],
\label{eq:sgal}
\end{equation}
with cyclostationary parameters $\mathcal{A}$, $f_1$, $f_2$, $f_{\rm knee}$, and $\alpha$ \cite[Table 2]{Digman2022}. The spectral fit is normalized to the $A, E, T$ TDI channels from Eq.~\eqref{eq:aetdef}. As the $T$ channel is null to leading-order in the long-wavelength limit, $X+Y+Z\simeq0$ and hence
\begin{equation}
0\simeq 3S_X+6S_c, \qquad S_c=-S_X/2,
\end{equation}
where $S_X$ and $S_c$ correspond to the self and cross-correlation channels in the $X, Y, Z$ basis respectively. Using Eq.~\eqref{eq:aetdef}, we also observe that $S_{AA}=S_{EE}=S_X-S_c= 3/2 \: S_X$. The foreground spectrum in the $X,Y,Z$ normalization is therefore
\begin{equation}
S_{\rm gal}^{XYZ}(f)
=
\frac{2}{3}S_{\rm gal}^{AE}(f).
\label{eq:aetoxyz}
\end{equation}

We next construct $\mathsf{C}_{\rm inst}(f)$ in the equal-arm limit from the self and cross-channel Michelson noise spectra. The total noise in a self and cross-correlation Michelson channels can be written in terms of the single-link optical metrology noise $P_{\rm OMS}(f)$ and test-mass acceleration noise $P_{\rm acc}(f)$ given in \cite[Eqs. 9 and 11]{babak2021} as
\begin{align}
P^{X_iX_i}(f)
&= \frac{1}{(2 L)^2} \left[
4 \, P_{\rm OMS}
+
8 \;\big(
1+\cos^2\chi\big)
\frac{P_{\rm acc}}{(2\pi f)^4}\right],
\label{eq:pn}
\\
P^{X_iX_j}(f)
&=
-\frac{\cos \chi}{(2L)^2}
\left[
2 \,P_{\rm OMS}
+
\frac{8 \,P_{\rm acc}}{(2\pi f)^4}
\right],
\label{eq:pncross}
\end{align}
where $\chi = f/f_\ast$ and $i \neq j$. The self-channel response has four contributions from the optical metrology noise and sixteen from the test mass acceleration noise \cite{Robson2019}, while the cross channels share one arm and have two contributions from the optical metrology noise and eight from the test mass acceleration noise, with a factor of $(-\cos \chi)$ set by the one-way offset between the two round trips. In the equal-arm limit, the TDI-1 channels in Eq.~\eqref{eq:X} have the prefactor $(1-\Dop^2)$ which scales the power spectrum as $|(1-\Dop^2)|^2 = 4\sin^2 \chi$. Thus, 
\begin{equation}
C_{\rm inst}^{X_iX_j}(f) = 4\chi^2\sin^2 \chi\,P^{X_iX_j}(f)
\label{eq:cinst}
\end{equation}
are the components of the instrumental covariance.
\begin{figure}
    \centering
    \includegraphics[width=\columnwidth]{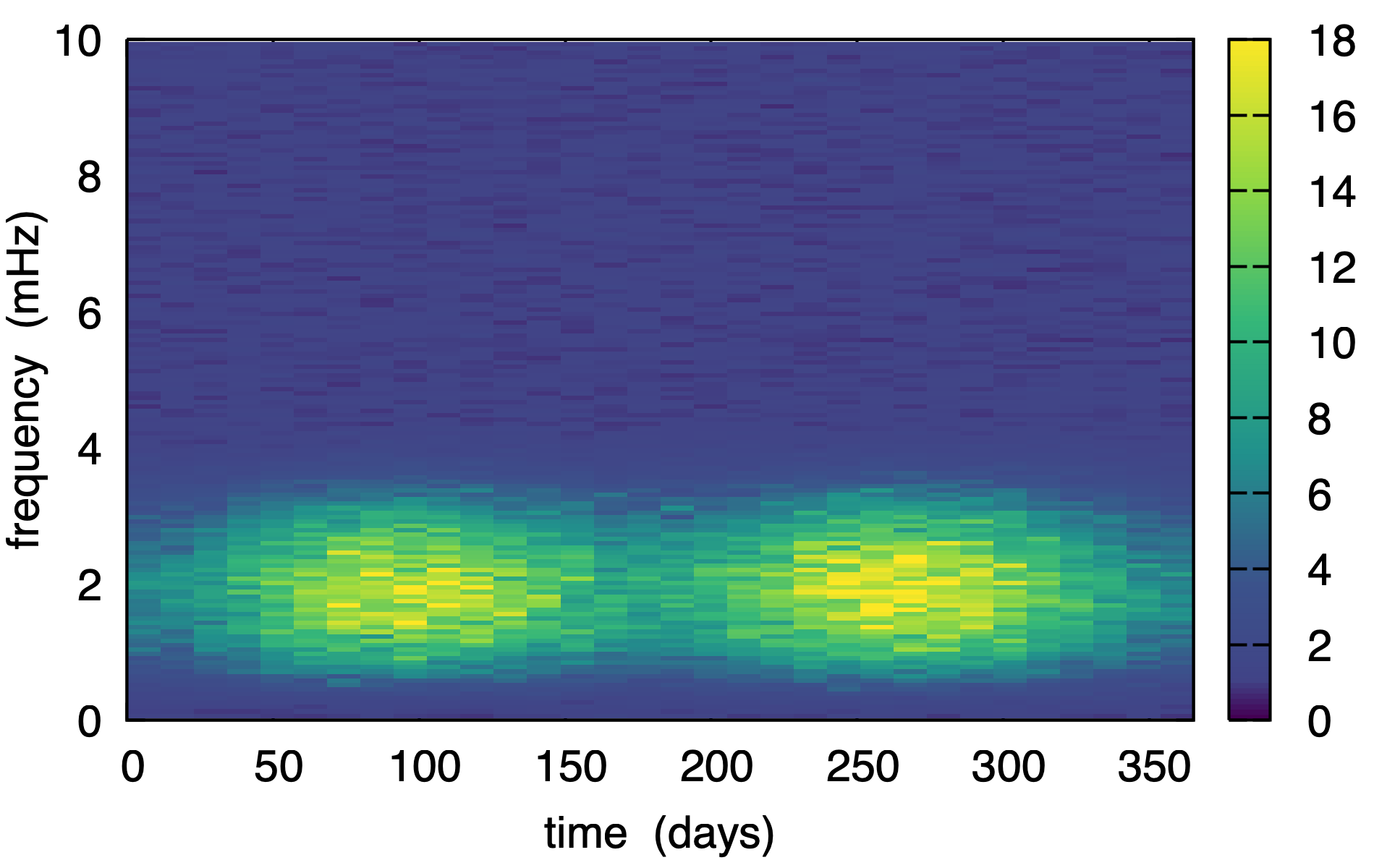}
    \caption{The WDM realization of one simulated year of LISA data, drawn from the full covariance $\mathsf{C}$ of Eq.~\eqref{eq:covsplit} in the $X$ channel, and whitened by the instrument noise to highlight the time and frequency structure of the annual Galactic modulation. Each column averages $N=128$ time pixels, or $11.4$ days. The spectral fit $S_\text{gal}(f)$ with cyclostationary parameters is adopted from \cite{Digman2022} and converted to the $X, Y, Z$ basis. The modulation matrix $\mathcal{M}$ is computed to leading-order for the fiducial Galaxy model described in Sec. \ref{subsec:healpix}. Instrument noise $\mathrm{C}_\text{inst}(f)$ is modeled using the single-link optical metrology noise $P_{\rm OMS}(f)$ and test-mass acceleration noise $P_{\rm acc}(f)$ from \cite[Eqs. 9 and 11]{babak2021}.}
    \label{fig:WDMspectrogram}
\end{figure}

We plot the WDM realization of one simulated year of LISA data in the $X$ channel, whitened by the instrument noise in order to highlight the contribution from the Galactic modulation in Fig.~\ref{fig:WDMspectrogram}. At each time-frequency pixel $(n,m)$ we draw the channel coefficients as
\begin{equation}
\bm{w}_{nm}=\mathsf{L}_{nm}\bm{z},
\qquad
\mathsf{C}_{nm}=\mathsf{L}_{nm}\mathsf{L}_{nm}^{\mathsf{T}},
\label{eq:wdmdraw}
\end{equation}
where $\mathsf{L}_{nm}$ is the lower triangular Cholesky factor of $\mathsf{C}_{nm}$ and $\bm{z}\sim\mathcal{N}(\bm{0},\mathsf{I})$. The $X$ channel reduces to $w_{X,nm}^2= C^{XX}(f_m, t_n)\,z_1^2$ with $C^{XX}(f_m, t_n)=C^{XX}_\text{inst}(f_m)+S_{\rm gal}(f_m)\mathcal{M}_{XX}(t_n)$, and whitening $w_{X}^2$ by the instrument noise gives,
\begin{equation}
\frac{w_{X,nm}^{2}}{C^{XX}_\text{inst}(f_m)}
=\left[1+\frac{S_{\rm gal}(f_m)\,\mathcal{M}_{XX}(t_n)}{C^{XX}_\text{inst}(f_m)}\right]z_1^2,
\label{eq:split}
\end{equation}
which is then averaged over $N=128$ time pixels. The foreground exceeds the instrument noise, $S_{\rm gal}\mathcal{M}_{XX}>C^{XX}_\text{inst}$, between about $0.3$ and $3.9\,\mathrm{mHz}$ and is cyclostationary as shown by the bright band in Fig.~\ref{fig:WDMspectrogram}.

\section{Results}\label{sec:Results}
\begin{figure}
  \includegraphics[width=\linewidth]{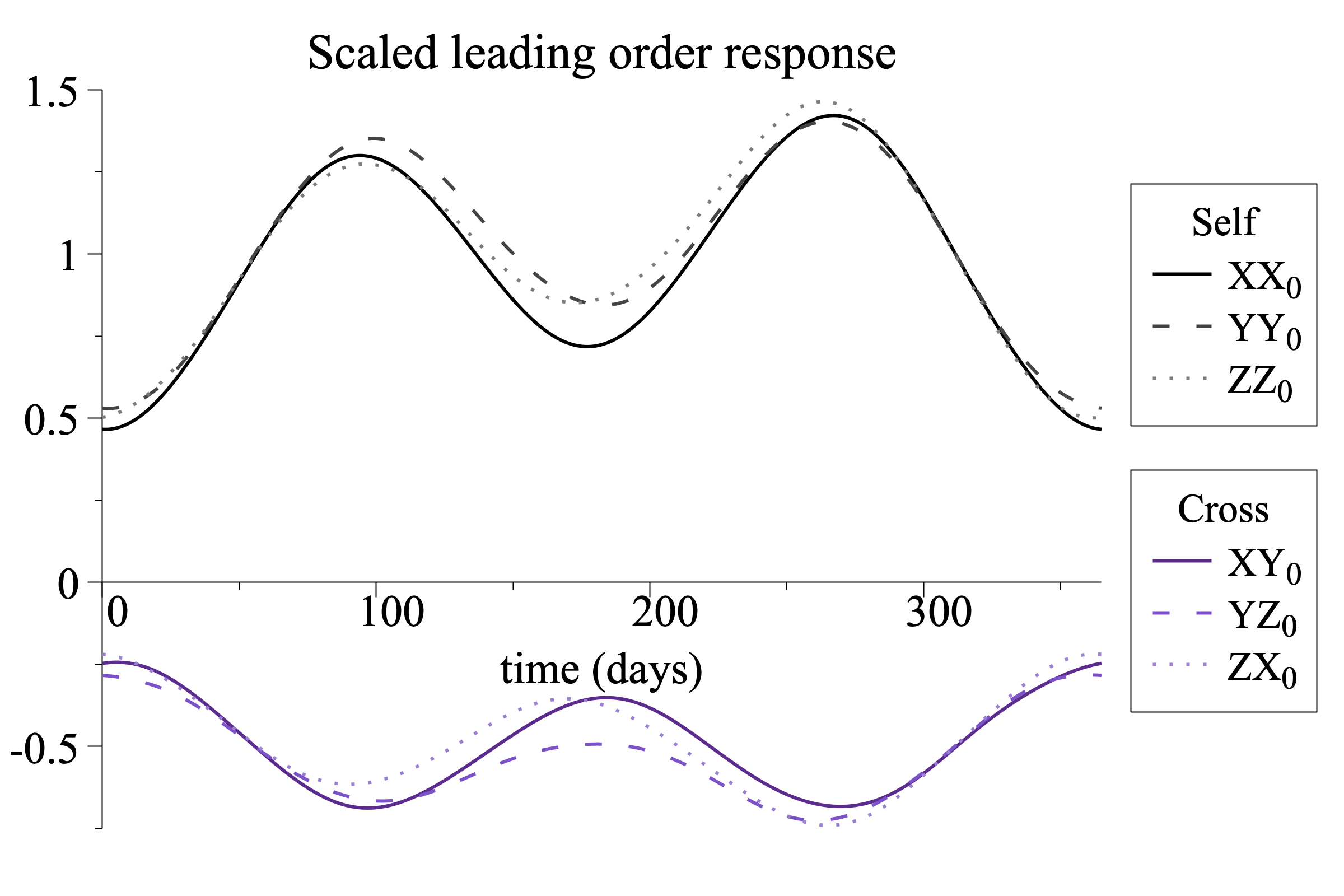}
  \caption{The scaled leading-order modulation of Galactic foreground in the TDI-1 XX, YY, ZZ (self-correlation) and XY, YZ, ZX (cross-correlation) channels. The modulation curves are normalized by the self-correlation channel annual mean.}
  \label{fig:lo_modcurves}
\end{figure}
We evaluate the modulation curves for the $X,Y,Z$ TDI-1 channels over one year of observation, with LISA arm length $L = 2.5\times10^{9}\,$m, transfer frequency $f_\ast = 1/2\pi L \simeq 19.1\,$mHz, and Keplerian orbital eccentricity $e = L/2\sqrt{3}R \simeq 0.0048$, which sets the amplitude of the arm-length variations $\epsilon_{ij}$. We adopt the two-component Galactic model with the fiducial values given in Sec.~\ref{subsec:healpix} and expand the response
coefficients as
\begin{equation}
  b_{\ell m}(f,t) \simeq b^{(0)}_{\ell m}(t) + e\,b^{\rm UA}_{\ell m}(t)
  + \left(\frac{f}{f_\ast}\right)^{2} b^{\rm FA}_{\ell m}(t),
  \label{eq:master_expansion}
\end{equation}
for each TDI-1 channel $X_iX_j$. Here $b^{(0)}_{\ell m}$ are the coefficients for the long-wavelength, equal-arm response, and $b^{\rm UA}_{\ell m}$ and $b^{\rm FA}_{\ell m}$ carry the unequal-arm and finite-arm corrections at
$\mathcal{O}(e)$ and $\mathcal{O}(f/f_\ast)^2$ respectively. As we discard the cross terms at $\mathcal{O}(e\,f/f_\ast)$ and higher, we can analyze the corrections independently. The leading-order modulation matrix $\mathcal{M}^{(0)}$ follows from Eq.~\eqref{eq:modmatrix} with $b^{(0)}_{\ell m}$ and is normalized by $\mathcal{N}^{(0)}$ as defined in Eq.~\eqref{eq:modmatrixNorm}. The corrected $\mathcal{M}$ follows from the corrected $b_{\ell m}$ and is re-normalized  by $\mathcal{N}$. We define the percentage correction to the leading-order response as
\begin{equation}
  \eta_{X_iX_j}(t,f) = 100\times\left|1-\frac{\mathcal{M}_{X_iX_j}(t,f)}
  {\mathcal{M}^{(0)}_{X_iX_j}(t)}\right|.
  \label{eq:percent_error}
\end{equation}
As both $\mathcal{M}$ and $\mathcal{M}^{(0)}$ are normalized by their respective self-correlation channel annual means, $\eta$ measures only the part of the correction that a constant rescaling cannot absorb. In the likelihood any such rescaling is absorbed by $S_{\rm gal}(f)$, which makes $\eta$ the physically relevant measure of the corrections.

\subsection{Leading-order response}\label{SubSec:LeadingOrder}

\begin{figure}
  \includegraphics[width=\columnwidth]{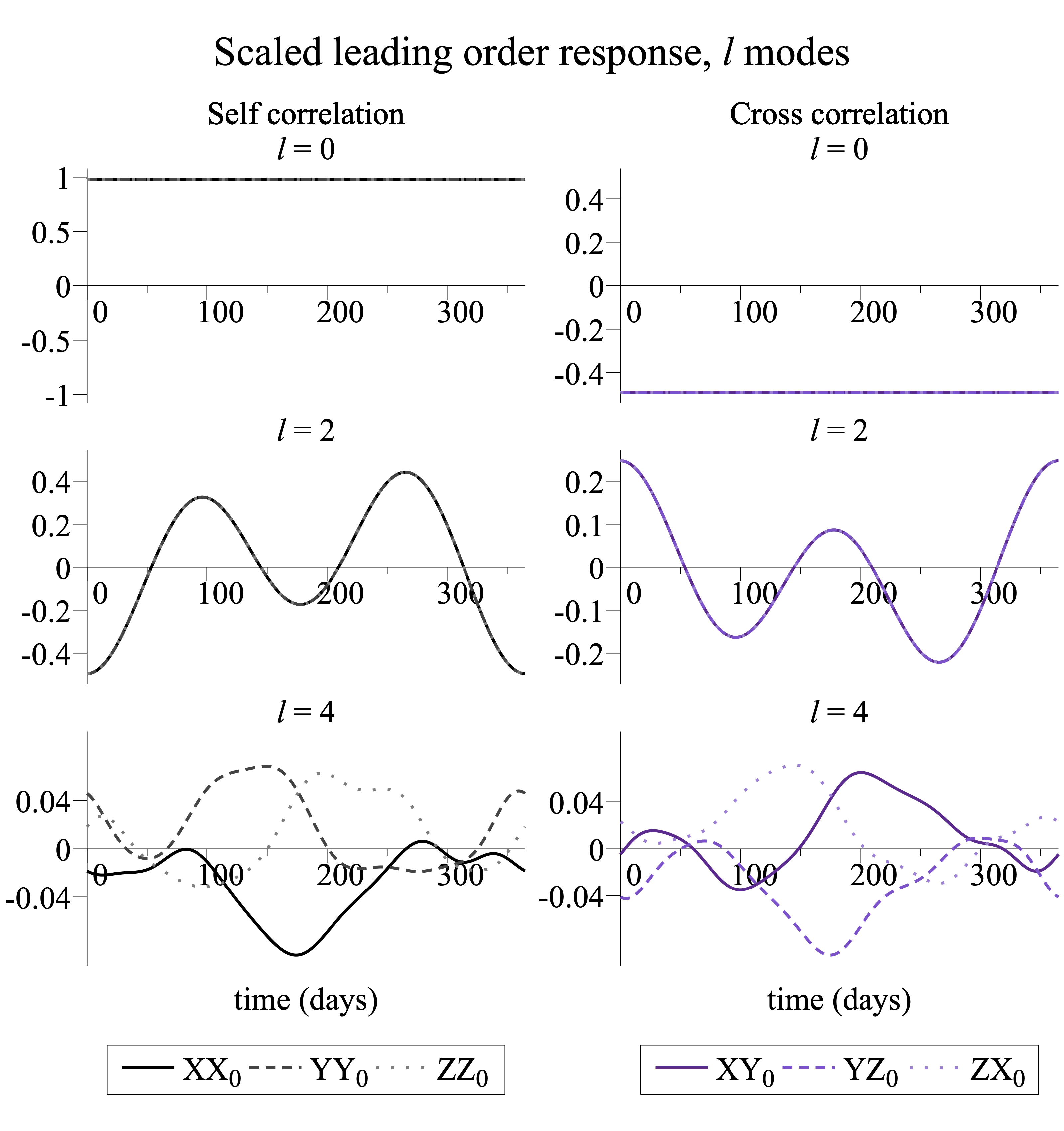}
  \caption{Multipole decomposition of the scaled leading-order response for the self (left) and the cross (right) correlation channels. The rows show the modulation curve contributions of the $\ell=0, 2, 4$ non-zero leading-order response modes respectively. The $\ell=0$ mode carries no time dependence, while the $\ell =2$ mode dominates the modulation for both the self and cross-correlation channels.}
  \label{fig:lo_lmodes}
\end{figure}

\begin{figure}
  \includegraphics[width=0.92\columnwidth]{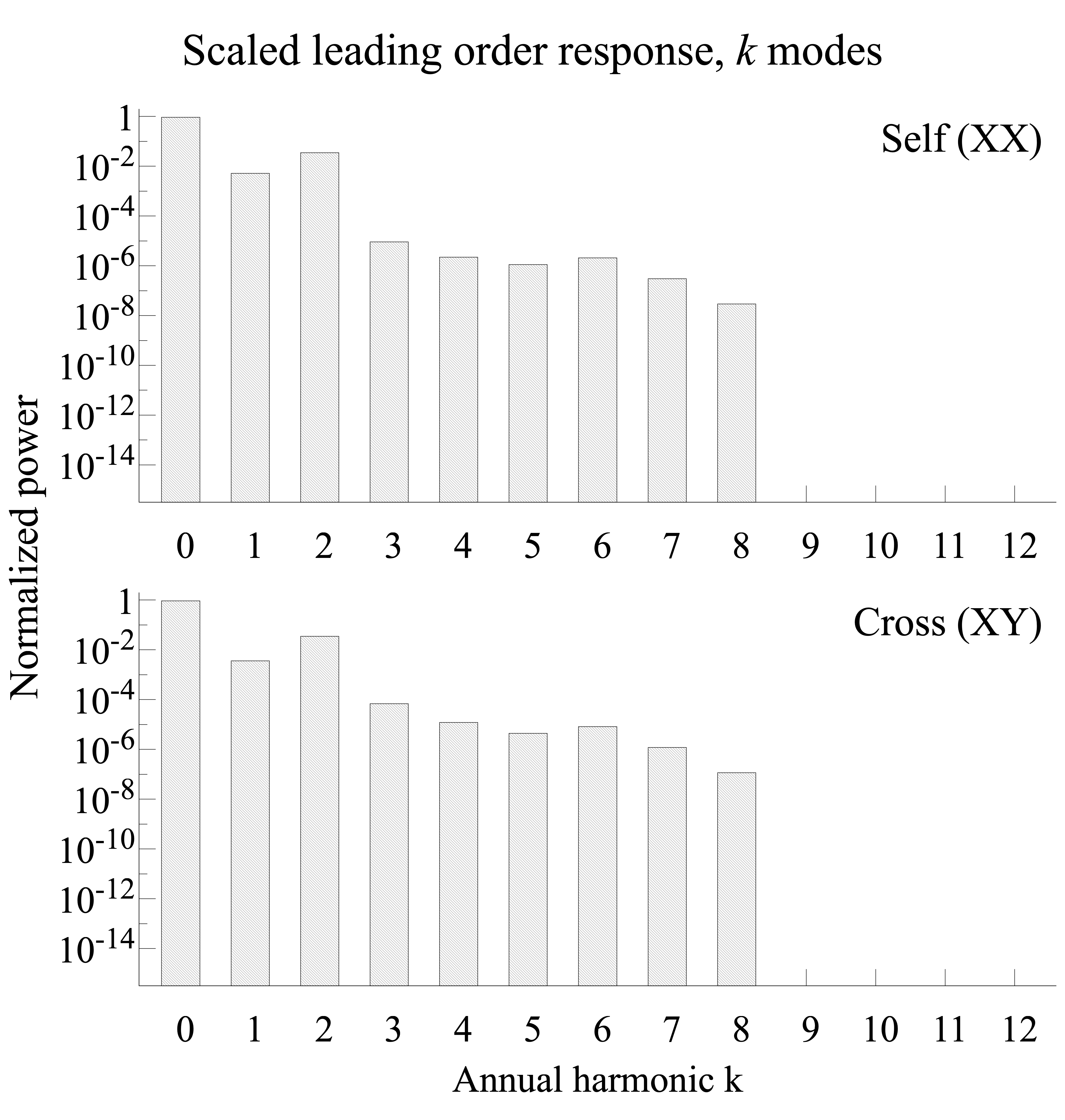}
  \caption{Normalized power in the annual modulation harmonics $k$ of the scaled leading-order response for the $XX$ (top) and the $XY$ (bottom) channels. The time dependence is introduced by the Wigner rotation on time-independent leading-order detector frame coefficients $\bar{b}_{\ell m}^{jk}$, distributing power over $|k|\leq 2 \ell_{\text{max}} = 8$ modes in the barycentric frame.}
  \label{fig:lo_harmonics}
\end{figure}

We plot the scaled leading-order modulation curves for self and cross-correlation channels in Fig.~\ref{fig:lo_modcurves}. The channels are cyclic under the rotation $X \to Y \to Z$. We decompose the scaled leading-order response into spherical harmonics and plot the non-zero $\ell=0,2,4$ modes for the scaled self and cross-correlation channels in Fig.~\ref{fig:lo_lmodes}. The $\ell=0$ mode carries no time dependence, while the $\ell =2$ mode dominates the modulation of the response. The leading-order response has no support above the $\ell =4$ mode.

We also plot the annual modulation harmonic decomposition of the $XX$ and $XY$ modulation curves. Expanding over a sidereal year, we highlight the fractional power $p^{ij}_{k}$ of the $k^{\text{th}}$ annual modulation harmonics in Fig.~\ref{fig:lo_harmonics} as 
\begin{equation}
    \mathcal{M}_{X_iX_j}(t) = \sum_{k} \mathcal{C}^{ij}_{k} e^{2\pi i k t/T_{\rm yr}}, \quad p^{ij}_{k} = \frac{|\mathcal{C}^{ij}_{k}|^{2}}{\sum_{k'}|\mathcal{C}^{ij}_{k'}|^{2}}
\end{equation}
Since the detector-frame coefficients $\bar{b}_{\ell m}^{jk}$ are time-independent, all of the time dependence of the modulation comes from the Wigner rotation. Thus, Eq.~\eqref{eq:wignerk} gives $|k| \le 2\ell_{\max} = 8$ in the barycentric frame.

\subsection{Finite armlength correction at \texorpdfstring{$\mathcal{O}(f/f_{*})^2$}{O(f/f*)}}
\begin{figure}[b]
  \includegraphics[width=\columnwidth]{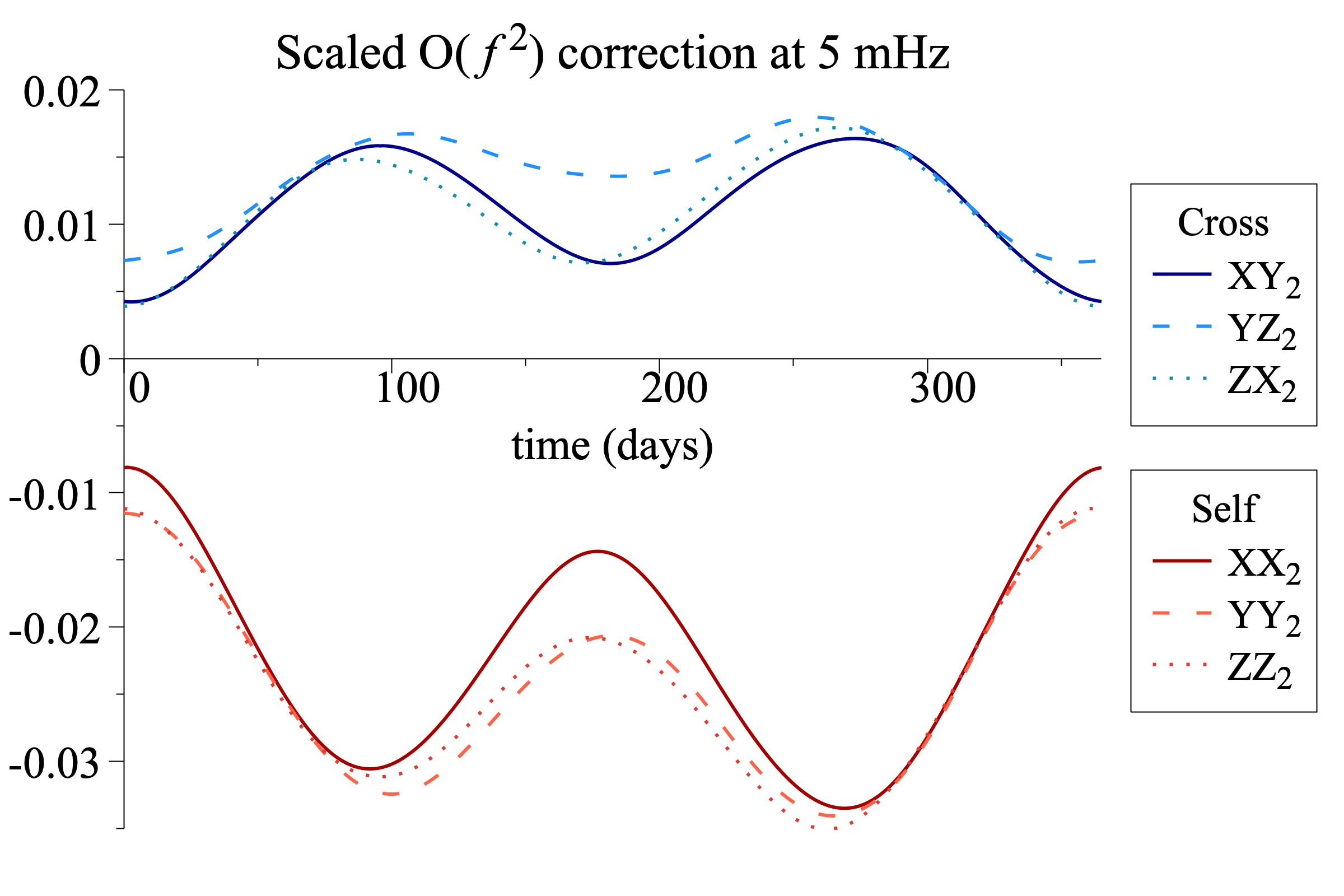}
  \caption{Scaled $\mathcal{O}(f/f_\ast)^2$ finite-arm corrections for the self and cross-correlation channels at $f = 5\,$mHz or $f/f_\ast \simeq 0.26$. The finite-arm correction changes both the magnitude and shape of the leading-order modulation curves.}
  \label{fig:fa_modcurves}
\end{figure}
\begin{figure}
  \includegraphics[width=\columnwidth]{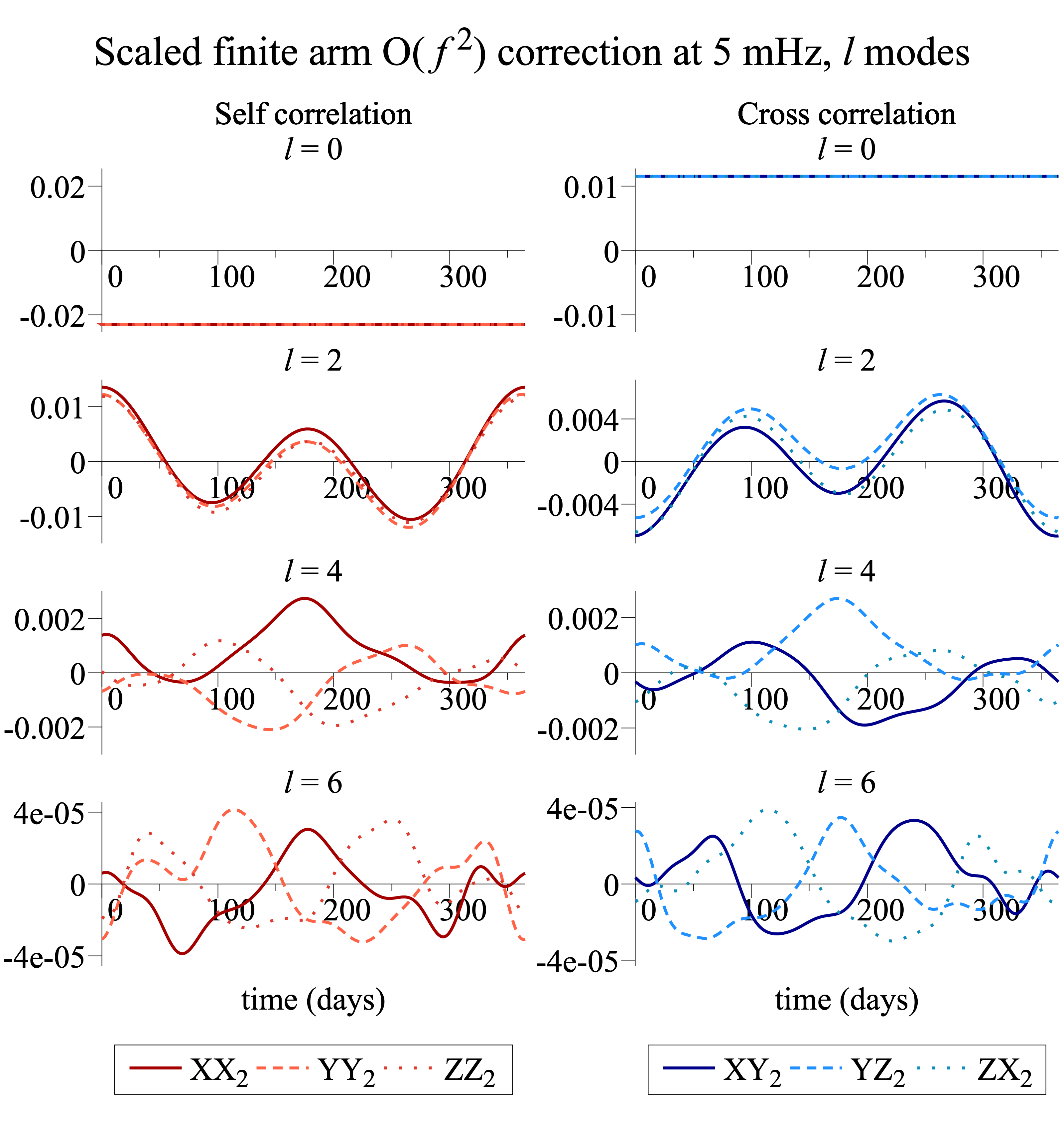}
  \caption{Multipole decomposition of the scaled finite-arm corrections for the self (left) and cross (right) correlation channels at $f = 5\,$mHz. The rows show the $\mathcal{O}(f/f_\ast)^2$ modulation corrections contributed by the $\ell=0, 2, 4, 6$ non-zero modes respectively. The finite armlength corrections change each $\ell$ mode and introduce higher-order angular structure at the $\ell_\text{max}=6$ mode.}
  \label{fig:fa_lmodes}
\end{figure}
\begin{figure*}  \includegraphics[width=0.78\textwidth]{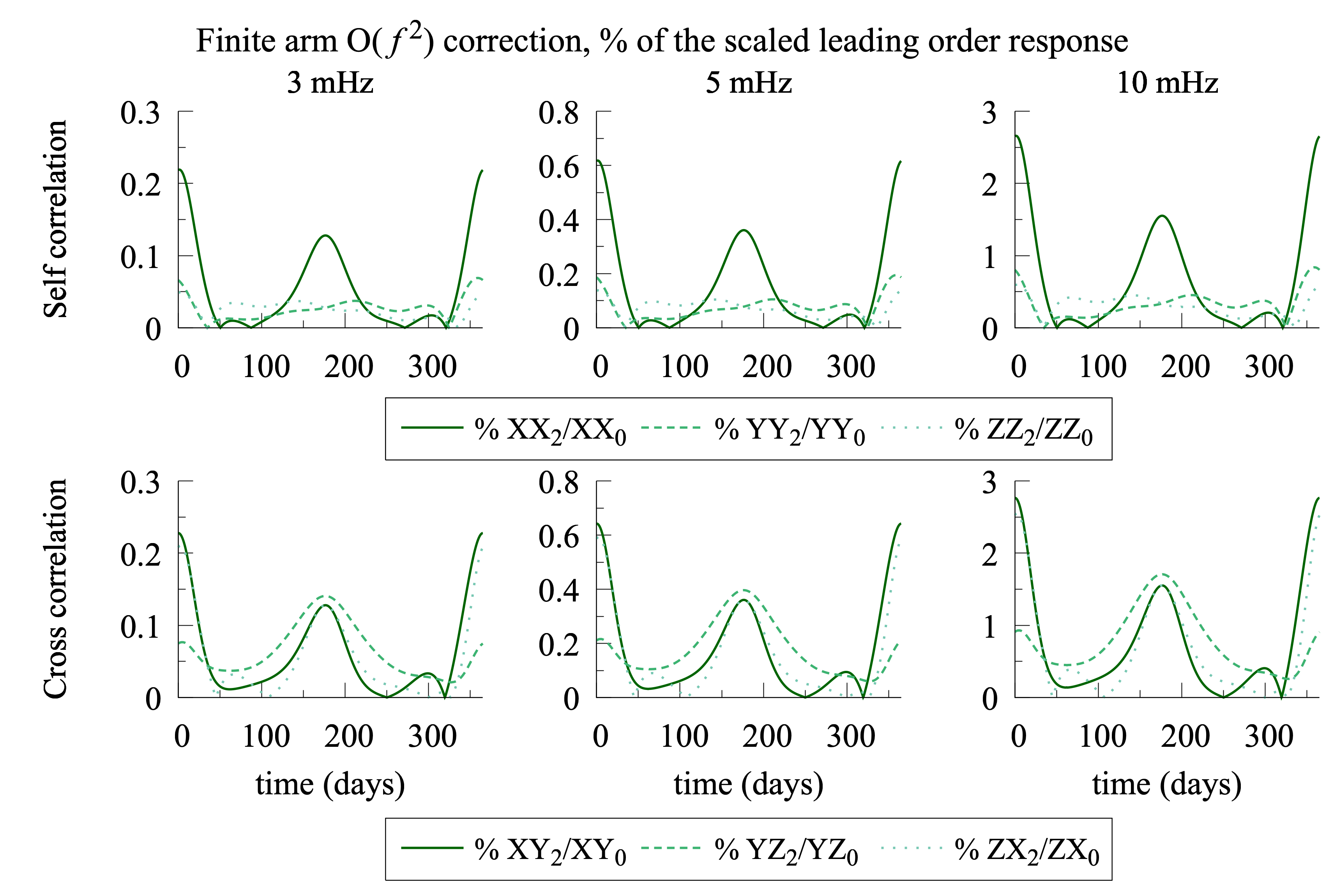}
  \caption{We plot the percentage finite-arm correction $\eta_{X_iX_j}^{FA}(t,f)$ to the scaled leading-order self (top row) and cross (bottom row) correlation channels at $f=3, 5$ and $10\,\text{mHz}$ (left to right). The correction varies strongly over the year and, for our fiducial Galaxy parameters, reaches a maximum value of $0.22\%$ ($0.23\%$) at $3\,\text{mHz}$, $0.62\%$ ($0.64\%$) at $5\,\text{mHz}$ and $2.67\%$ ($2.77\%$) at $10\,\text{mHz}$ in the self (cross) correlation channels.}
  \label{fig:fa_percent_corr}
\end{figure*}
\begin{figure}
  \includegraphics[width=\columnwidth]{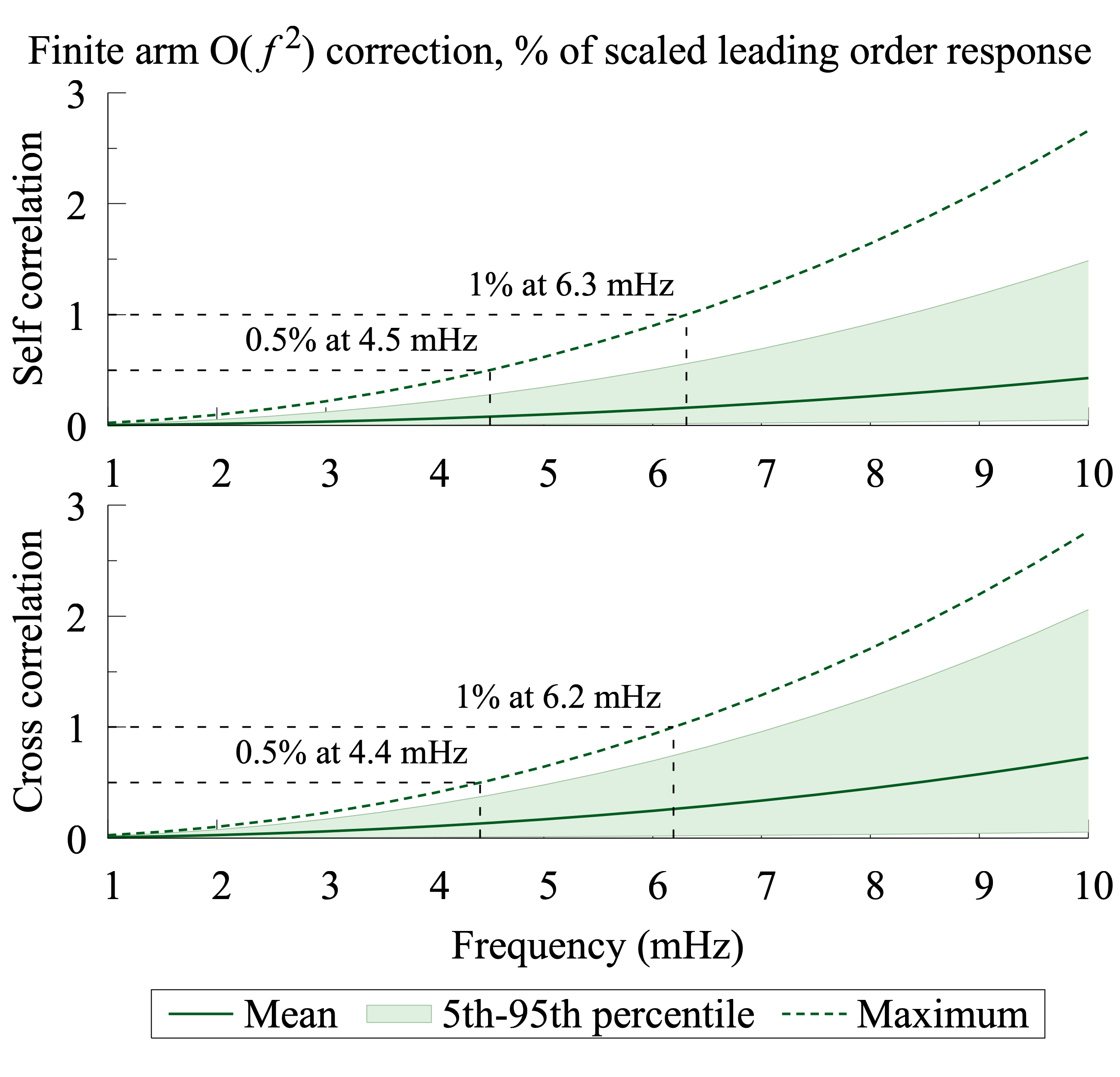}
  \caption{Percentage finite-arm correction as a function of frequency from $1$ to $10\,$mHz for the self (top) and cross (bottom) correlation channels. The band spans variation of the correction over the year and across channels, while the central line plots the year and channel-averaged value. Dashed lines indicate when the maximum correction crosses the $0.5\%$ and $1\%$ levels for our fiducial Galaxy parameters.}
  \label{fig:fa_band}
\end{figure}
\begin{figure}
  \includegraphics[width=0.92\columnwidth]{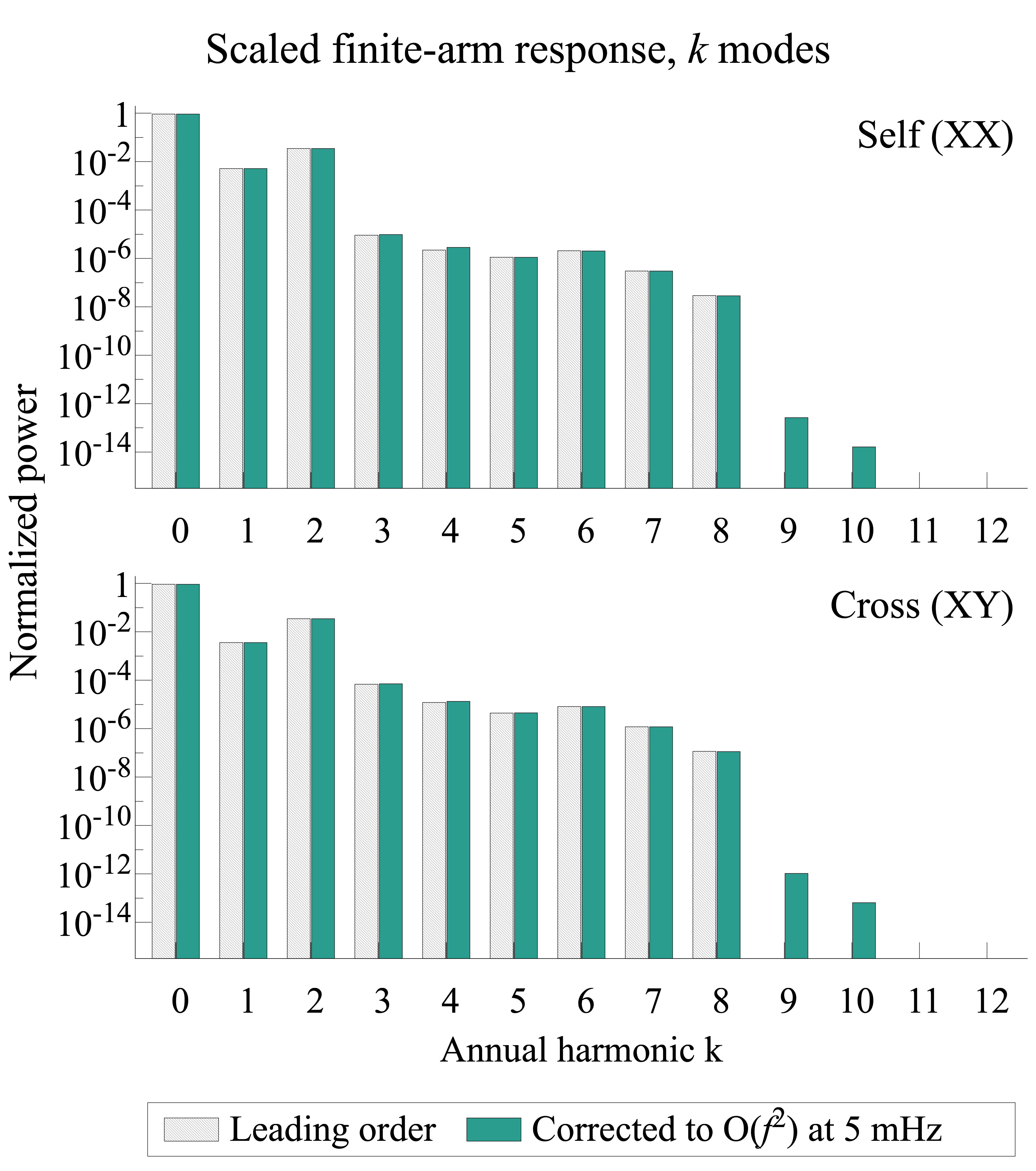}
  \caption{Normalized power in the annual modulation harmonics $k$ of the scaled leading-order and finite-arm corrected $XX$ (top) and $XY$ (bottom) channel responses at $f=5 \,\text{mHz}$ in the barycentric frame.}
  \label{fig:fa_harmonics}
\end{figure}

The finite-arm corrections, which relax the long-wavelength approximation, enter the modulation curves at the second order in frequency. We plot the scaled $\mathcal{O}(f/f_\ast)^2$ finite-arm corrections for the self and cross-correlation channels at $f = 5 \,\text{mHz}$, which corresponds to $f/f_{*} \simeq 0.26$, in Fig.~\ref{fig:fa_modcurves}.
The finite-arm correction changes both the magnitude and shape of the modulation curves. The detector frame seed $\bar{b}_{\ell m}^{jk}$ gains corrections to every mode already present at the leading-order and acquires support for additional modes with $(l,m) = (2, \pm2), (4,\pm2), (6, 0), (6,\pm2)$ and $(6,\pm4)$ at $\mathcal{O}(f/f_\ast)^2$. The Wigner rotation then distributes power over $m \in[-\ell,\ell]$ for $l=0,2,4,6$. Thus, with finite-arm corrections, angular resolution of the response improves with higher $\ell_{\text{max}} = 6$.  We plot the contribution of each $\ell$ mode to the finite-arm self and cross channel corrections at $f=5\, \text{mHz}$ in Fig.~\ref{fig:fa_lmodes}. We also plot the percentage finite-arm correction $\eta_{X_iX_j}^{FA}(t,f)$ to the scaled leading-order self and cross-correlation channels at $f=3, 5$ and $10 \;\text{mHz}$, corresponding to $f/f_\ast \simeq 0.16,\, 0.26$ and $0.52$, in Fig.~\ref{fig:fa_percent_corr}. At $10 \;\text{mHz}$, the Galactic binary population is largely resolved and we plot the correction only for demonstration. The finite-arm correction grows quadratically with frequency and varies strongly over the year. Table~\ref{tab:finite_arm} lists the channel and year averaged correction, along with the maximum correction over the year, at 3, 5, and $10\,$mHz for the fiducial Galaxy parameters given in Sec.~\ref{subsec:healpix}. We highlight the percentage finite-arm correction as a function of frequency from $1$ to $10\;$mHz for the self and cross-correlation channels in Fig.~\ref{fig:fa_band}. The band spans the variation of the correction over the year and across channels, while the central line plots the year and channel-averaged value. For our fiducial Galaxy parameters, the maximum correction in self (cross) correlation channel crosses the $0.5\%$ level at $f = 4.5\,$mHz ($4.4\,\text{mHz}$) and the $1\%$ level at $f = 6.3\,$mHz ($6.2\,\text{mHz}$). In Fig.~\ref{fig:fa_harmonics}, we compare the annual modulation harmonic content for the scaled leading-order and finite-arm corrected responses at $f = 5\, \text{mHz}$ in the barycentric frame.
\begin{table}
\begin{ruledtabular}
\begin{tabular}{ccccc}
 & \multicolumn{2}{c}{Averaged} & \multicolumn{2}{c}{Maximum} \\
\cline{2-3}\cline{4-5}
$f$ (mHz) & Self & Cross & Self & Cross \\
\hline
3  & $0.035$ & $0.06$ & $0.22$ & $0.23$ \\
5  & $0.10$  & $0.17$  & $0.62$ & $0.64$ \\
10 & $0.42$  & $0.72$  & $2.67$ & $2.77$ \\
\end{tabular}
\caption{\label{tab:finite_arm} Percentage finite-arm corrections to the scaled leading-order self and cross-correlation channels for the fiducial Galaxy parameters in Sec.~\ref{subsec:healpix}. The finite-arm correction grows quadratically with frequency and varies strongly over the year. The first pair of columns contain the channel and year averaged corrections, while the second pair contain the maximum corrections over the year at 3, 5, and $10\,$mHz.}
\end{ruledtabular}
\end{table}

\subsection{Unequal-arm correction at \texorpdfstring{$\mathcal{O}(e)$}{O(epsilon)}}\label{SubSec:UnequalArm}
\begin{figure}[b]
  \includegraphics[width=\columnwidth]{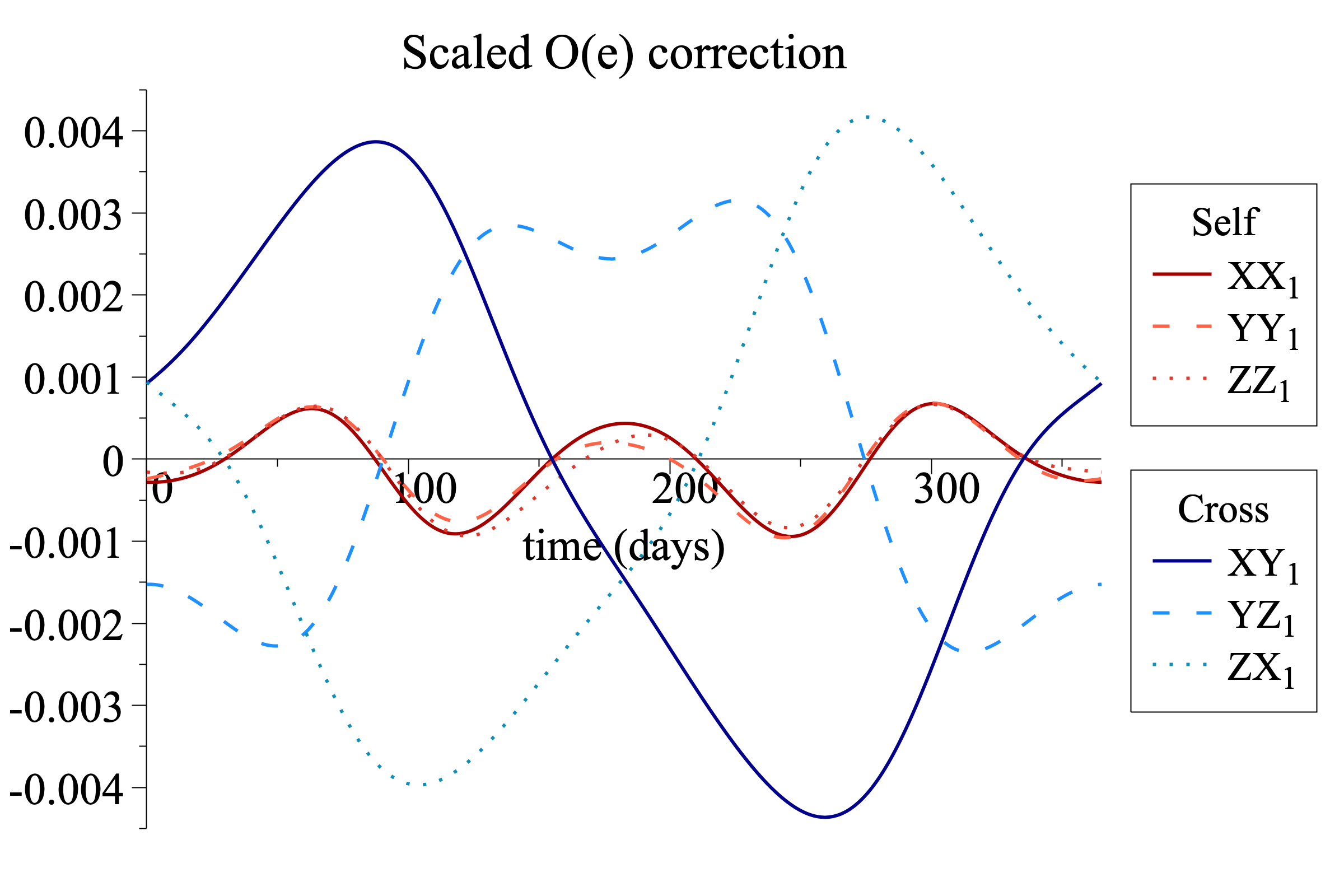}
  \caption{Scaled $\mathcal{O}(e)$ unequal-arm corrections for the self and cross-correlation channels at Keplerian orbital eccentricity $e \simeq 0.0048$. The unequal-arm correction changes both the magnitude and shape of the modulation curves.}
  \label{fig:ua_modcurves}
\end{figure}

\begin{figure}
  \includegraphics[width=\columnwidth]{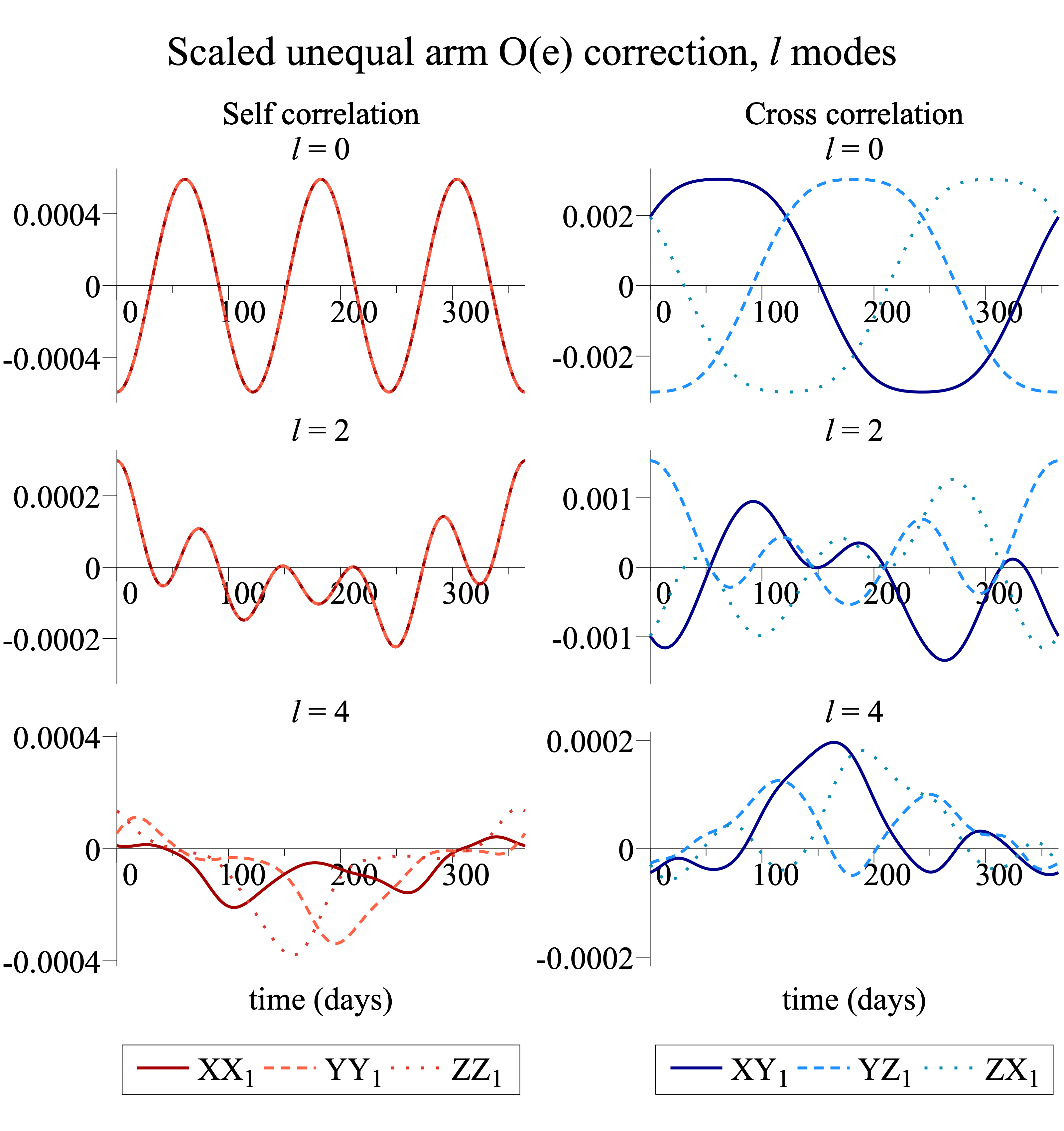}
  \caption{Multipole decomposition of the scaled unequal-arm corrections for the self (left) and the cross (right) correlation channels. The rows show the $\mathcal{O}(e)$ modulation corrections contributed by the $\ell=0, 2, 4$ non-zero modes respectively. The unequal-arm corrections do not introduce any higher-order angular structure, preserving the $\ell_\text{max}=4$ extent of the leading-order response.}
  \label{fig:ua_lmodes}
\end{figure}

We relax the equal-arm assumption for Keplerian spacecraft orbits up to the first order in orbital eccentricity. We plot the scaled $\mathcal{O}(e)$ unequal-arm corrections for the self and cross-correlation channels in Fig.~\ref{fig:ua_modcurves}. Unequal-arm corrections contain the same $\ell, m$ modes as the leading-order response, but change both the magnitude and shape of the response. Wigner rotation to the barycentric frame preserves these non-zero $\ell = 0,2,4$ modes of the correction. We plot the contribution of each of these $\ell$ modes to the unequal-arm self and cross-correlation channel corrections in Fig.~\ref{fig:ua_lmodes}.

\begin{figure}
\includegraphics[width=0.92\columnwidth]{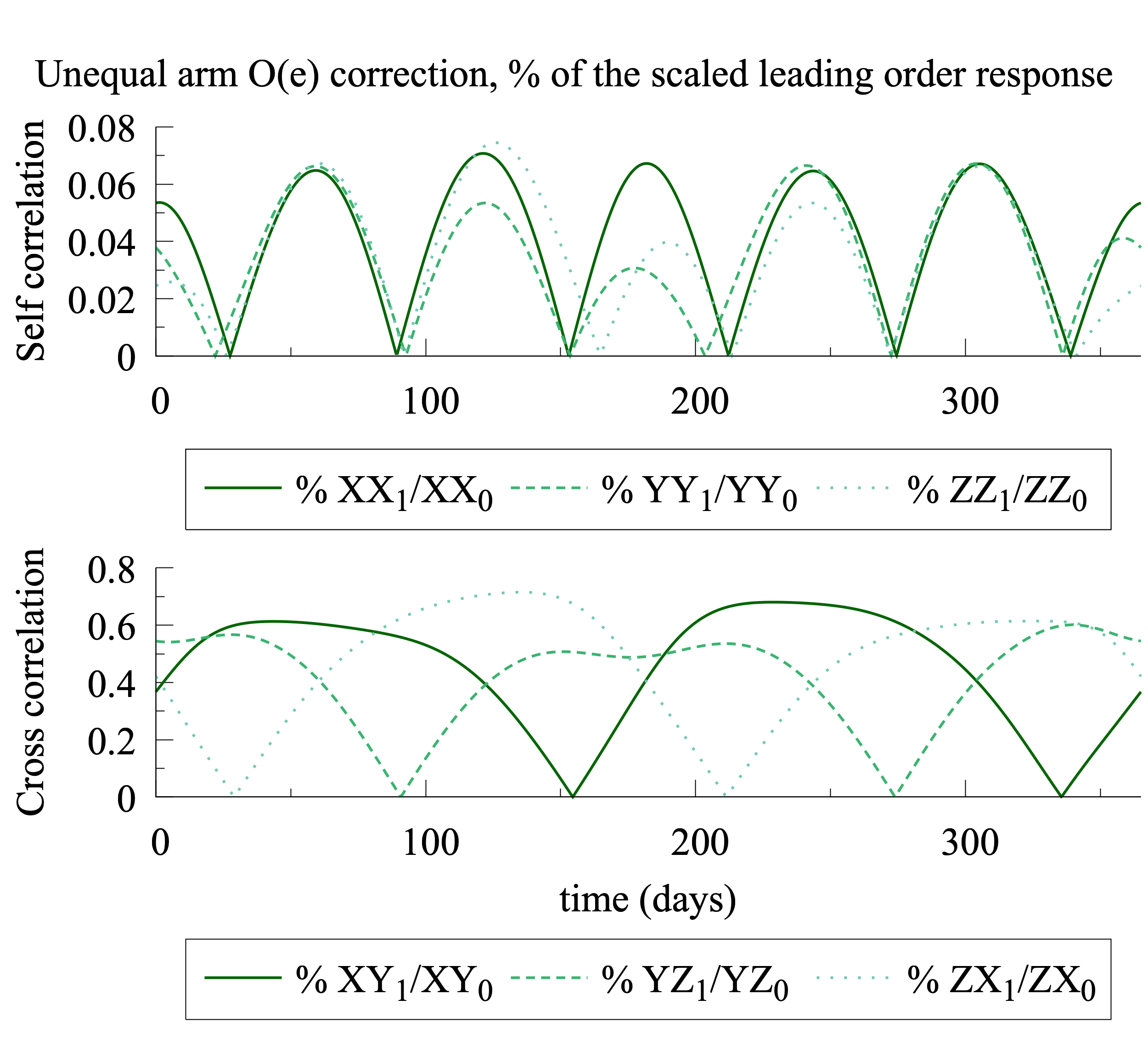}
  \caption{Percentage unequal-arm correction $\eta_{X_iX_j}^{UA}(t,f)$ to the scaled leading-order self (top) and cross (bottom) correlation channels. The channel and year averaged unequal-arm correction for the self (cross) correlation channel is $0.037\%$ ($0.44\%$) with our fiducial Galaxy parameters, while the maximum value of the correction is $0.074 \%$ ($0.71\%$) over the year.}
  \label{fig:ua_percent_corr}
\end{figure}

\begin{figure}
  \includegraphics[width=0.92\columnwidth]{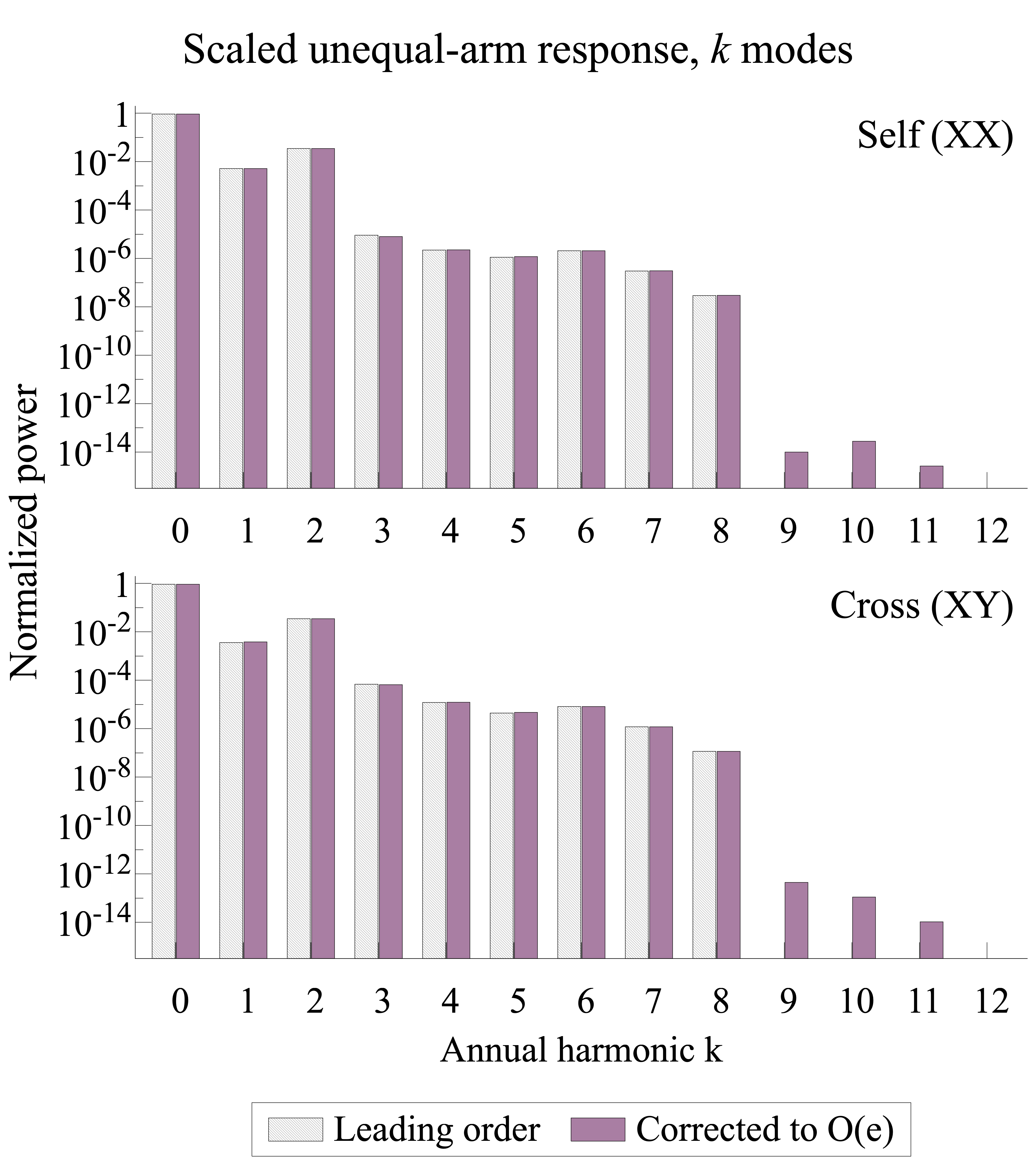}
  \caption{Normalized power in the annual modulation harmonic $k$ of the scaled leading-order and unequal-arm corrected $XX$ (top) and $XY$ (bottom) channel responses in the barycentric frame. Wigner rotation to the barycentric frame on time-dependent detector frame seeds $\bar{b}_{\ell m}^{jk}$ with $\bar{k}=\pm1, \pm3$ modes for Keplerian spacecraft orbits at $\mathcal{O}(e)$, shifts the annual harmonic indices $k$ by $\bar{k}$, extending the range of the response’s annual modulation harmonic content.}
  \label{fig:ua_harmonics}
\end{figure}

We also plot the percentage unequal-arm correction $\eta_{X_iX_j}^{UA}(t,f)$ to the scaled leading-order self and cross-correlation channels in Fig.~\ref{fig:ua_percent_corr}. The correction strongly varies over the year. The channel and year averaged unequal-arm correction for the self (cross) correlation channel is $0.037\%$ ($0.44\%$), while the maximum value of the correction reaches $0.074 \%$ ($0.71\%$) over the year for our fiducial Galaxy parameters. Unlike the finite-arm corrections, the unequal-arm corrections are frequency-independent. Comparing the year and channel-averaged values of the unequal-arm correction with those of the finite-arm correction in Fig.~\ref{fig:fa_band}, calculated for the fiducial Galaxy parameters, we observe that the mean unequal-arm correction dominates the mean finite-arm correction below $f = 3.1\,$mHz ($7.9\,$mHz), while its maximum value exceeds the finite-arm correction maximum value below $f = 1.7\,$mHz ($5.2\,$mHz) for the self (cross) correlation channels. We also compare the annual modulation harmonic content for the scaled leading-order and unequal-arm corrected responses in Fig.~\ref{fig:ua_harmonics}. The unequal-arm correction terms make the detector frame coefficients $\bar{b}_{\ell m}^{jk}$ time-dependent, with $\bar{k}=\pm1, \pm3$ modes for Keplerian spacecraft orbits. The Wigner rotation to the barycentric frame acting on these time-dependent coefficients then shifts the annual harmonic indices $k$ by $\bar{k}$. This extends the range of the response’s annual modulation harmonic content from $|k| \leq 8$ for the leading-order response to $|k| \leq 11$ for the response with $\mathcal{O}(e)$ unequal-arm corrections.

\subsection{Dependence on the Galactic foreground model}\label{SubSec:GalaxyDependence}
\begin{figure}[b]
  \includegraphics[width=\columnwidth]{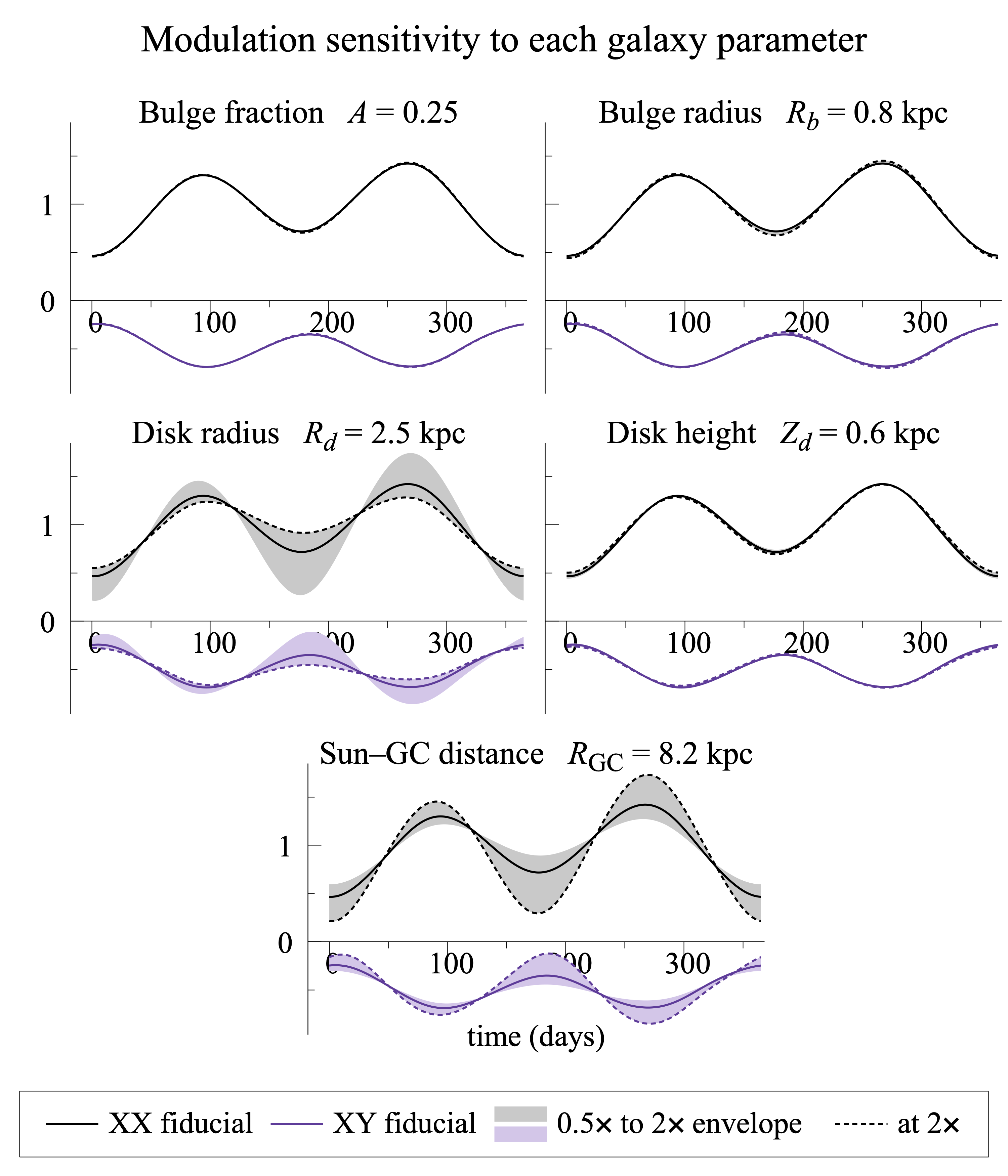}
  \caption{Change in the leading-order modulation curves $XX$ and $XY$ as each Galactic parameter is varied between half and twice its fiducial value. Solid (dashed) lines show the modulation curves at the fiducial (doubled) parameter values, while the shaded bands highlight the $0.5\times$ to $2\times$ envelope. Increasing $A$ or $R_{b}$ concentrates the emission towards the Galactic center and increasing $R_{\rm GC}$ accentuates anisotropy, deepening the modulation. Increasing $R_{d}$ or $Z_{d}$ disperses the emission and reduces the modulation depth.}
  \label{fig:galaxy_curves}
\end{figure}

The Galactic model affects the leading-order response as well as the finite and unequal-arm corrections. We plot the leading-order $XX$ and $XY$ modulation curves with each Galactic parameter doubled and halved about its fiducial value in Fig.~\ref{fig:galaxy_curves}. Increasing the bulge fraction $A$ or the bulge radius $R_{b}$ concentrates the emission density toward the Galactic center and deepens the peaks and troughs of the modulation curve. Increasing the disk width $R_{d}$ or the disk height $Z_{d}$ spreads the emission density across a larger solid angle and reduces the peak-trough contrast of the modulation curves. Increasing $R_{\rm GC}$ places the center of the Galaxy farther away, deepens the peak-trough contrast, and has the opposite effect of increasing $R_{d}$ and $Z_{d}$. A common rescaling of $R_{b}$, $R_{d}$, $Z_{d}$ and $R_{\rm GC}$ is degenerate and leaves the angular distribution and modulation curves nearly unchanged. Thus, a compact Galaxy with a nearby center and a proportionally larger, distant Galaxy produce the same modulation.

\begin{figure*}
  \includegraphics[width=\textwidth]{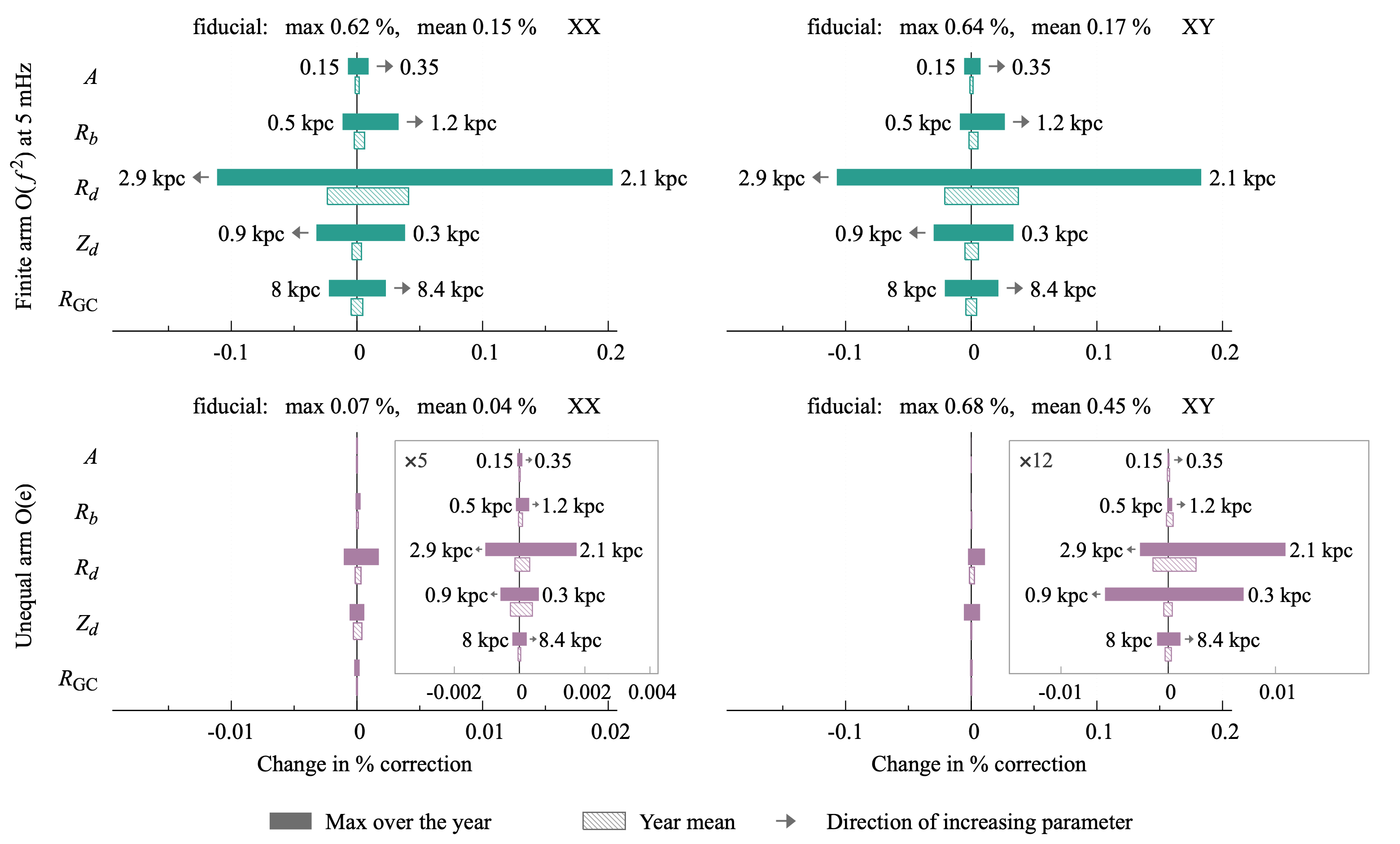}
  \caption{Change in the finite (row 1) and unequal-arm (row 2) percentage corrections for the $XX$ (left) and $XY$ (right) scaled leading-order response as each Galactic parameter is varied over astrophysically motivated intervals around our fiducial parameter values. The vertical zero line marks percentage corrections to the scaled leading-order $XX$ and $XY$ response for our fiducial Galaxy parameters. Solid (hatched) bars highlight the change in the annual maximum (mean) percentage correction. The arrows next to the bar indicate the direction of increasing Galactic parameter value. The unequal-arm percentage corrections for the $XX$ and $XY$ channels are also plotted at $\times \,5$ and $\times \,12$ magnification of the $x$-axis respectively. The finite-arm corrections show a greater variation with changes in Galactic parameters, indicating greater sensitivity to anisotropy, while the unequal-arm corrections remain relatively independent of Galactic parameter variations.}
  \label{fig:galaxy_corrections}
\end{figure*}

We highlight the variation in finite and unequal-arm corrections with changing Galactic parameters in Fig.~\ref{fig:galaxy_corrections}. We vary each parameter over an astrophysically motivated range, and quantify the uncertainty in the finite and unequal-arm corrections that follows from the uncertainty in the Galactic model. We span $R_{d} \in [2.1, 2.9]$~kpc from \cite{BlandHawthorn2016}, $Z_{d} \in [0.3, 0.9]$~kpc (thin to thick disk) from \cite{Juric2008, BlandHawthorn2016}, $R_{\rm GC} \in [8.0, 8.4]$~kpc from \cite{GRAVITY2019, BlandHawthorn2016, McMillan2011}, $R_{b} \in [0.5, 1.2]$~kpc from \cite{Sofue2009} and $A \in [0.15, 0.35]$, $\pm 0.1$ about the fiducial value from \cite{Adams2014}. Varying the Galactic parameters changes the anisotropy of the Galactic distribution and, consequently, the angular structure of the modulation curves. The finite-arm correction shows a greater dependence on Galactic anisotropy as compared to the unequal-arm case. As the anisotropy increases with increasing $A$, $R_b$ or $R_{\text{GC}}$, higher-$\ell$ modes expose additional angular structure in the finite-arm correction.  Conversely, as the anisotropy decreases with increasing $R_d$ or $Z_d$, the flat $\ell=0$ mode of the finite-arm correction (Fig.~\ref{fig:fa_lmodes}) contributes more prominently. The correction then has reduced time-dependent structure and a larger fraction of it is therefore absorbed into the overall magnitude rescaling. The unequal-arm correction behaves differently as its $\ell=0$ mode retains time-dependent structure from the varying arm lengths (Fig.~\ref{fig:ua_lmodes}). This structure changes the shape of the modulation and cannot be removed by a constant rescaling, so decreasing the Galactic anisotropy produces a smaller change in the unequal-arm correction. Moreover, while the finite-arm correction extends to higher multipoles, the unequal-arm correction has the same $\ell_{\max}$ as the leading-order response. It is therefore less sensitive to the angular structure introduced as the Galactic anisotropy increases. Note that bars show the effect over the astrophysically allowed interval, not the intrinsic sensitivity to a fixed fractional change in each parameter. Thus, the large $R_d$ dependent variation partly reflects its broader allowed range rather than a uniquely strong parameter dependence.

\subsection{Fisher forecast for the Galactic parameters}\label{SubSec:Fisher}
\begin{figure}
  \includegraphics[width=0.49\textwidth]{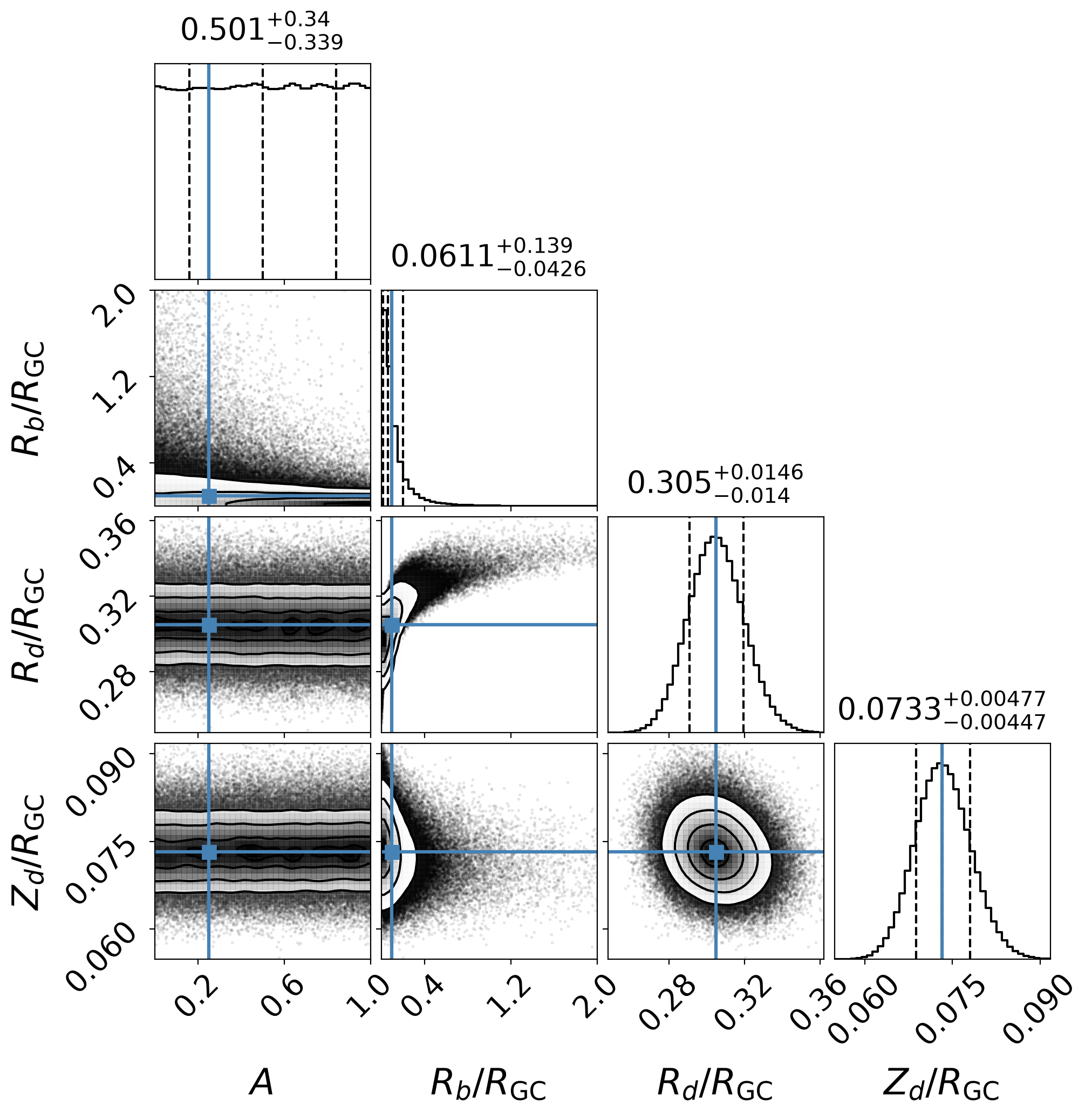}
  \caption{Parameter distributions sampled from the Fisher matrix-based Gaussian approximation ${\cal N}(\bm{q}_0',\Gamma^{-1})$ for one year of the Galactic foreground modulation, where $\bm{q}' = \left(A,\,\ln\!\left(R_b/R_{\rm GC}\right),\, \ln\!\left(R_d/R_{\rm GC}\right),\, \ln\!\left(Z_d/R_{\rm GC}\right) \right)$ and accepted samples are subject to the physical bound $A \in [0,1]$. Distances are rescaled relative to $R_{\rm GC}$ to remove the overall scale degeneracy. Blue lines are the injected values. The disk scale width $R_d$ and disk scale height $Z_d$ relative to $R_{\rm GC}$ are recovered to the $7.8\%$ and $12\%$ levels respectively (Table~\ref{tab:fisher_sigma}). The bulge fraction $A$ recovers its uniform prior on $[0,1]$. Imposing the physical prior range on $A$ yields a bound on the bulge radius.}
  \label{fig:corner}
\end{figure}

We evaluate the Fisher matrix from Eq.~\eqref{eq:fisher} for a one year observation in the long-wavelength, equal-arm limit, where the modulation matrix $\mathcal{M}_{X_iX_j}^{(0)} (f,t)$ contains only the leading-order response. As the modulation is invariant under a common rescaling of Galactic length scales, we re-parametrize distances relative to $R_{\rm GC}$ and work in the projected basis 
\begin{equation}
    \bm{q}' = \left(A,\,\ln\!\left(R_b/R_{\rm GC}\right),\, \ln\!\left(R_d/R_{\rm GC}\right),\, \ln\!\left(Z_d/R_{\rm GC}\right) \right),
\end{equation} 
injecting the re-parametrized fiducial Galactic parameters $\bm{q}_0' = \left(0.25,\,\ln\!\left(0.0976\right),\, \ln\!\left(0.3049\right),\, \ln\!\left(0.0732\right) \right)$. We sample from the Gaussian approximation ${\cal N}\,(\bm{q}_0',\,\Gamma^{-1})$, impose the physical bound $A \in [0,1]$ by sample rejection, and plot resulting parameter distributions in Fig.~\ref{fig:corner}. The diagonal of the inverse Fisher matrix gives the marginalized one-sigma uncertainty on each parameter $q^a$ after marginalizing over the remaining parameters
\begin{equation}
\sigma_a = \sqrt{\left(\Gamma^{-1}\right)_{aa}} ,
\label{eq:sigma}
\end{equation}
\begin{table}[b]
\centering
\begin{tabular}{@{}l@{\hskip 2.0em}c@{}}
\hline
\textbf{Parameter} $q^a$ & \bm{$\sigma_a$} \\
\hline
$A$                      & 23.18  \\
$\ln(R_b/R_{\rm GC})$    & 43.43  \\
$\ln(R_d/R_{\rm GC})$    & 0.0778 \\
$\ln(Z_d/R_{\rm GC})$    & 0.1200 \\
\hline
\end{tabular}
\caption{Marginalized Fisher forecast uncertainties $\sigma_a$ for the fiducial Galaxy model in Sec. \ref{subsec:healpix} at $f_{\max}=10~{\rm mHz}$. Normalized modulation carries information about the disk scale width $R_d$ and disk scale height $Z_d$ relative to $R_{\rm GC}$, at the $7.8\%$ and $12\%$ levels respectively. By contrast, the bulge fraction $A$ and bulge radius $R_b$ are strongly correlated.}
\label{tab:fisher_sigma}
\end{table}
which is the Cram\'er-Rao bound for minimum uncertainty of any unbiased estimator \cite{TegmarkTaylorHeavens1997}. In Table~\ref{tab:fisher_sigma}, we report the Fisher forecast uncertainties for the fiducial Galaxy model of Sec.~\ref{subsec:healpix} with $f_{\max}=10~{\rm mHz}$. The disk geometry is well measured. We recover $R_{d}/R_{\rm GC} = 0.305^{+0.015}_{-0.014}$ and $Z_{d}/R_{\rm GC} = 0.0733^{+0.0048}_{-0.0045}$ (Fig.~\ref{fig:corner}) with fractional precisions of $7.8\%$ and $12.0\%$ (Table~\ref{tab:fisher_sigma}). Both medians coincide with their injected values. The bulge fraction and radius are individually unconstrained but strongly correlated, $\rho_{A,\ln R_{b}} = -0.9997$. The modulation of the unresolved Galactic foreground therefore constrains the shape of the disk while leaving the absolute scale degenerate with $R_{\rm GC}$.

\section{Conclusion and Discussion}\label{sec:Conclusion}
We develop a fast, forward model to map arbitrary parameterized Galaxy models with integrable luminosity densities to the cyclostationary modulation of the Galactic foreground in the LISA channels, by separating the Galactic model from the detector response. The Galactic luminosity density is numerically integrated and decomposed into spherical harmonics. The spherical harmonic coefficients of the LISA antenna pattern are precomputed in the detector frame, corrected for finite-arm effects up to second order in frequency and unequal-arm effects up to first order in orbital eccentricity, and rotated to the barycentric frame using Wigner D matrices. At leading-order the response occupies only \(\ell=0,2,4\) modes in the detector frame. Rotating to the barycentric frame then distributes the power over all $|m| \leq \ell$, but preserves the $\ell$ modes. Unequal-arm effects follow the same angular structure as the leading-order response. Finite-arm effects add $\mathcal{O}(f/f_\ast)^2$ corrections to every leading-order mode in both self and cross-spectra, and populate the additional modes $(\ell,m)=(2,\pm2)$, $(4,\pm2)$, $(6, 0)$, $(6,\pm2)$, and $(6,\pm4)$ in the detector frame. In the cross-spectra, inter-vertex phases produce purely imaginary odd-parity modes, namely $(3,\pm3)$ and $(5,\pm3)$ at first order in $f/f_*$, which do not affect the modulation curves, but should be accounted for in the quadrature spectrum and likelihood evaluations.

Unequal-arm corrections are frequency-independent at $\mathcal{O}(e)$ and can reach an annual maximum of $\sim 0.074\%$ ($0.71\%$) in the self (cross) correlation channels, whereas finite-arm corrections can reach an annual maximum of $\sim 0.22\% \,(0.23\%)$  at 3 mHz and $\sim 0.62\% \,(0.64\%)$ at 5 mHz in the self (cross) correlation channels, with the exact values depending upon the Galaxy parameters and the anisotropy of the Galaxy. The  channel and year averaged mean unequal-arm correction dominates the mean finite-arm correction below $f \simeq 3.1\,$mHz ($7.9\,$mHz), while its maximum value exceeds the finite-arm correction maximum value below $f \simeq 1.7\,$mHz ($5.2\,$mHz) for the self (cross) correlation channels.

We apply our forward model in a Fisher forecast for a one-year observation, treating the bulge fraction and the four Galactic length scales as free parameters. In the WDM basis, the likelihood factorizes over time-frequency pixels, and the Galactic parameters enter solely through the normalized modulation matrix of the cyclostationary foreground. As this modulation depends only on the angular distribution of the foreground, the unresolved binaries carry no information about the absolute distance scales. A compact Galaxy with a nearby center and a proportionally larger, distant Galaxy produce the same modulation. The Fisher matrix is therefore nearly blind to a common rescaling of all lengths, and we re-parametrize to remove the overall scale degeneracy. The disk geometry is well determined, with $R_{d}/R_{\rm GC}$ and $Z_{d}/R_{\rm GC}$ recovered to $7.8\%$ and $12.0\%$ levels respectively. The bulge sector is degenerate, with the bulge fraction and radius anti-correlated at $\rho = -0.9997$ and individually unconstrained. Resolved binaries complement this purely angular information and can provide the absolute length scale needed to break the overall scale degeneracy, as the distance from the Sun to a detached binary can be determined from its measured frequency, frequency derivative, and amplitude \cite{Adams2012,Korol2019}. A joint analysis of the resolved and unresolved populations would thus fix the absolute Galactic scale and sharpen the constraints obtained here from the unresolved foreground.

\section{Future Directions and Scope}\label{sec:future}

The framework we develop here can be extended to incorporate arm flexing, adopt more realistic Galactic source models, and potentially perform joint inference on the resolved and unresolved Galactic binary populations. The unequal-arm corrections derived here assume constant arm lengths over the light-crossing time. Accounting for flexing arms in real data requires time-dependent delays and a second-generation TDI description \cite{Shaddock2003,TintoDhurandhar2021}. Flexing also modifies the orthogonality of the $A$, $E$, and $T$ channels and makes the factor relating second-generation variables to their first-generation counterparts channel-dependent \cite{Hartwig2023}. The framework also readily accommodates more realistic source distributions. The harmonic factorization allows us to compute the modulation for a new Galactic model by recomputing only the source coefficients, while the detector coefficients remain unchanged. This makes it straightforward to include structures such as the bar, spiral arms, and thick disk. Our framework can also be applied to spatially distinct populations such as the Magellanic Clouds, the stellar halo, and extragalactic sources. The computational efficiency of the forward model makes a full posterior analysis of the Galactic hyperparameters feasible, extending inference beyond the local forecast provided by the Fisher matrix. A joint likelihood for the resolved binary catalog and the unresolved foreground modulation would allow the unresolved majority of the source population to constrain Galactic parameters, as well as refine the foreground noise model that enters the global fit and limits low-frequency parameter estimation. Requiring a common set of Galactic parameters would also test whether a single density model describes the resolved and unresolved Galactic binary populations consistently.

\vspace{\baselineskip}
\begin{acknowledgments}
This work was supported by NASA LISA Preparatory Science Grant 80NSSC19K0320. The authors thank Dr. Tyson B. Littenberg and Dr. Robert J. Rosati for their feedback and insights.
\end{acknowledgments}
\bibliography{bibliography} 
\end{document}